\documentclass[notitlepage,12pt]{jedm}\usepackage[]{graphicx}\usepackage[]{color}
\makeatletter
\def\maxwidth{ %
  \ifdim\Gin@nat@width>\linewidth
    \linewidth
  \else
    \Gin@nat@width
  \fi
}
\makeatother

\definecolor{fgcolor}{rgb}{0.345, 0.345, 0.345}

\usepackage{framed}
\makeatletter
 {\par\unskip\endMakeFramed%
 \at@end@of@kframe}
\makeatother

\definecolor{shadecolor}{rgb}{.97, .97, .97}
\definecolor{messagecolor}{rgb}{0, 0, 0}
\definecolor{warningcolor}{rgb}{1, 0, 1}
\definecolor{errorcolor}{rgb}{1, 0, 0}
\usepackage{alltt}

\usepackage{subcaption}
\usepackage{fullpage}
\usepackage{enumerate}
\usepackage{multirow}
\usepackage{comment}
\usepackage{amsmath,amssymb,amsfonts,amsthm}
\usepackage{bbm}
\usepackage{setspace}
\usepackage{verbatim}
\usepackage{bm}
\usepackage{pdflscape}
\usepackage{tikz}
\usepackage{xr}
\usepackage{makecell}

\title{Mastery Learning in Practice:\\
A (Mostly) Descriptive Analysis of Log Data from the Cognitive Tutor Algebra I Effectiveness Trial}

\author{{\large Anita Israni}\\University of Texas College of
  Education\\aisrani@gmail.com \and {\large Adam C Sales}\\University of Texas College of
  Education\\asales@utexas.edu  \and {\large John F Pane}\\RAND Corporation\\jpane@rand.org}

\date{}

\IfFileExists{upquote.sty}{\usepackage{upquote}}{}
\begin{document}
\maketitle

\begin{abstract}
Mastery learning, the notion that students learn best if they move on
from studying a topic only after having demonstrated mastery, sits at
the foundation of the theory of intelligent tutoring. This paper is an
exploration of how mastery learning plays out in practice, based on
log data from a large randomized effectiveness trial of the Cognitive
Tutor Algebra I (CTAI) curriculum. We find that students frequently
progressed from CTAI sections they were working on without
demonstrating mastery and worked units out of order. Moreover, these
behaviors were substantially more common in the second year of the
study, in which the CTAI effect was significantly larger. We explore
the various ways students departed from the official CTAI curriculum,
focusing on heterogeneity between years, states, schools, and
students. The paper concludes with an observational study of the
effect on post-test scores of teachers reassigning students out of
their current sections before they mastered the requisite skills,
finding that reassignment appears to lowers posttest scores---a finding
that is fairly reseliant to confouding from omitted
covariates---but that the effect varies substantially between
classrooms.
\end{abstract}
\section{Introduction}
Mastery learning sits at the foundation of intelligent tutoring
systems \cite{corbett2001cognitive,wenger2014artificial}.
The philosophy of mastery learning assumes a well-structured
curriculum, and posits that students progress within the curriculum as
they master its skills
\cite{bloom1968learning,kulik1990effectiveness}. The Cognitive Tutor
Algebra I (CTAI) system, developed by Carnegie Learning, Inc., is one of the best studied and best regarded examples of modern educational software. It is a blended learning system for teaching algebraic concepts and principles, to middle and high school students, including both textbook materials and software. The software component of the curriculum allows students to progress at their own pace and receive individualized feedback on their performance. A large-scale randomized effectiveness trial conducted by the RAND corporation showed that, in some circumstances, CTAI boosts students' scores on an Algebra-I posttest by about one fifth of a standard deviation \cite{pane2014effectiveness}. CTAI's success in this experiment would seem to validate its pedagogy: mastery learning, and the algebra I curriculum on which it is based.

However, the theory underlying CTAI does not always determine its
use. To be sure, the software has a standard set of algebra topics,
divided into units and further into sections; and a standard sequence
for presenting them. But this precise curriculum is not mandatory. At
the request of educators, it can be customized by altering what units
or sections are included (including, possibly, material from a
different standard curriculum such as geometry), as well as their
sequence, to conform to local or state standards or scope and sequence
guides. Further, teachers have the option of moving students within
the curriculum, regardless of the software's estimate of their skill
mastery.

This article examines teachers' and schools' adherence and
non-adherence to the standard, mastery-based CTAI curriculum, using
data from the RAND study, a seven-state randomized controlled trial of
CTAI in high schools and middle schools. That study found a significant
positive effect of CTAI in high schools, during their second year of
implementation but not the first. Students in the treatment group of
the study were enrolled in one or more of the standard or customized
curricula during their participation in the study, and the software
logged aspects of their usage, including time spent, sections
encountered, and whether the software judged the students to have
mastered the sections. Adopting standard procedures of an
effectiveness trial, both Carnegie Learning and the researchers
running the study restricted their support and oversight to what is
typically provided outside of an experimental context. Thus, the
software data from the study reflect typical usage. Secondary data analyses used principal stratification to show that students who attempted more sections experienced larger treatment effects, and students who had high or low assistance levels, as opposed to an average level, experienced smaller treatment effects \cite{sales2015exploring,sales2016student}.
\citeN{sales2017role} found that students more likely to master worked
sections of the CTAI software may experience \emph{smaller} effects
than those less likely to achieve mastery, casting doubt on the role
of mastery learning as a mechanism for the treatment effect.

Here we contextualize the previous findings to describe, in detail, the ways
in which schools, teachers, and students violate mastery learning.
We will begin with short discussions of the RAND effectiveness trial and the usage
data it produced.
Sections \ref{sec:curricula}---\ref{sec:order} will describe overall
patterns of usage.
First, we will discuss standard and customized CTAI curricula used in
the study (Section \ref{sec:curricula}).
Next, in Section \ref{sec:usage}, we will describe patterns in the amount CTAI usage, showing that
it varied widely between states, between years, between schools, and between students.
In particular, the amount of usage decreased from years 1 to
2---though more in some states than others.
Then in section \ref{sec:order} we will describe how the amount of usage
changed---which units of the CTAI curriculum were worked more and less
from years 1 to 2---and find that order in which students worked
CTAI's units varied across years. We show that this change is due
mostly, but not completely, to the presence of customized curricula in
year 2.

In Section \ref{sec:mastery} we will describe patterns of mastery in general,
and in Section \ref{sec:cp} we will delve deeply into ``reassignment'': the process in which a
teacher moves a student out of a section he or she has not (yet)
mastered into a new section.
In particular, we will attempt to elucidate teachers' goals in
reassigning students. One hypothesis is the need for teachers to push
ahead students who were falling behind, i.e. to reassign them from
sections on which they were struggling, to allow them catch up with
the rest of the class. Another hypothesis is that teachers sought to
push students past easier sections to begin working on more relevant
or challenging topics for them. A third hypothesis is that teachers needed to cover certain topics in preparation for an upcoming state exam, and might have reassigned groups of students all at the same time to cover topics that might otherwise not have been covered. This hypothesis may lead to an increase in reassignments as the exam approaches.
We find evidence of all three motivations for reassignment, varying
across classrooms, schools, states, and study  years.

Finally, in Section \ref{sec:effects} will will give
quasi-experimental estimates of the effects of reassignment on
students' posttest scores.
We find that reassignment probably decreases student learning,
though the effects vary widely across classrooms.
Section \ref{sec:discussion} will conclude the article with a summary
of findings and discussion.

\section{The RAND Effectiveness Trial}\label{sec:RANDtrial}
The study to measure the effectiveness of CTAI included 7 states, 73
high schools, and 74 middle schools with nearly 18,700 high school
students and 6,800 middle school students participating. Schools were enrolled in a
total of 52 school districts that were distributed among urban,
suburban, and rural areas. Schools were matched on a set of
covariates, and then randomly assigned to the treatment or control
group. Schools in the control group continued with their current
algebra curriculum, and schools in the treatment group used Carnegie
Learning's curriculum which includes CTAI textbook materials and sofware. Each school participated for two years, with a different cohort
of students participating the second year (with a small fraction of
students present in the study both years because they repeated
algebra). It should be noted that this study did not include statewide
implementations; the study results cannot be generalized to all
schools within the state. In some states, one large school district
participated, while in other states, a set of smaller school districts
participated. The states included Alabama (AL), Connecticut (CT),
Kentucky (KY), Louisiana (LA), Michigan (MI), New Jersey (NJ), and
Texas (TX). Each state participated in both the middle school and high
school arms of the study, except AL, which participated only in the
middle school arm. The current study focuses on high school students
only.

There are some limitations to the available data for this study.
Log data from some schools, and some students within schools, were
missing either because the log files were not retrievable, or because
of an imperfect ability to link log data to other study data files.
For this reason, this study uses only data from the
18 treatment schools for which at least
80\%
of students in both study years appear in the log data file.
This sample includes 4460 students, around
75\%
of the treated high-school sample.
Table \ref{tab:nByState} gives the number of students in the sample by
state and year.
The states in the table are ordered by the total number of students
they represent in the sample; they will appear in this order in
all of the forthcoming tables and figures.
Some figures will only show data from a subset of states; since so few
students were in New Jersey, it will be excluded from almost all
state-by-state comparisons (but included analyses that pool across states).
% latex table generated in R 3.3.1 by xtable 1.8-2 package
% Fri Dec 15 13:54:41 2017
\begin{table}[ht]
\centering
\begin{tabular}{rrrrrrr}
  \hline
 & TX & KY & MI & LA & CT & NJ \\
  \hline
Year 1 & 947 & 890 & 325 & 164 & 112 &  17 \\
  Year 2 & 719 & 646 & 285 & 180 & 139 &  36 \\
   \hline
\end{tabular}
\caption{Numbers of students in the sample by state and study year}
\label{tab:nByState}
\end{table}

In this sample from 18 schools, 164
students who participated in the RAND study do not appear in the log data; they may have not used the CTAI
software at all, or may have been excluded from the log data for other
reasons.
Since we don't know which is true, we exclude these students from most
analyses.

It is likely that some usage data were missing, even for students who
appear in the usage dataset.
However, it is impossible to know in which cases these data were
missing or why; for the most part, we ignore this problem, but it
should be kept in mind nonetheless.

\section{Standard and Customized Curricula}\label{sec:curricula}

\begin{figure}
  \centering

\includegraphics[width=\maxwidth]{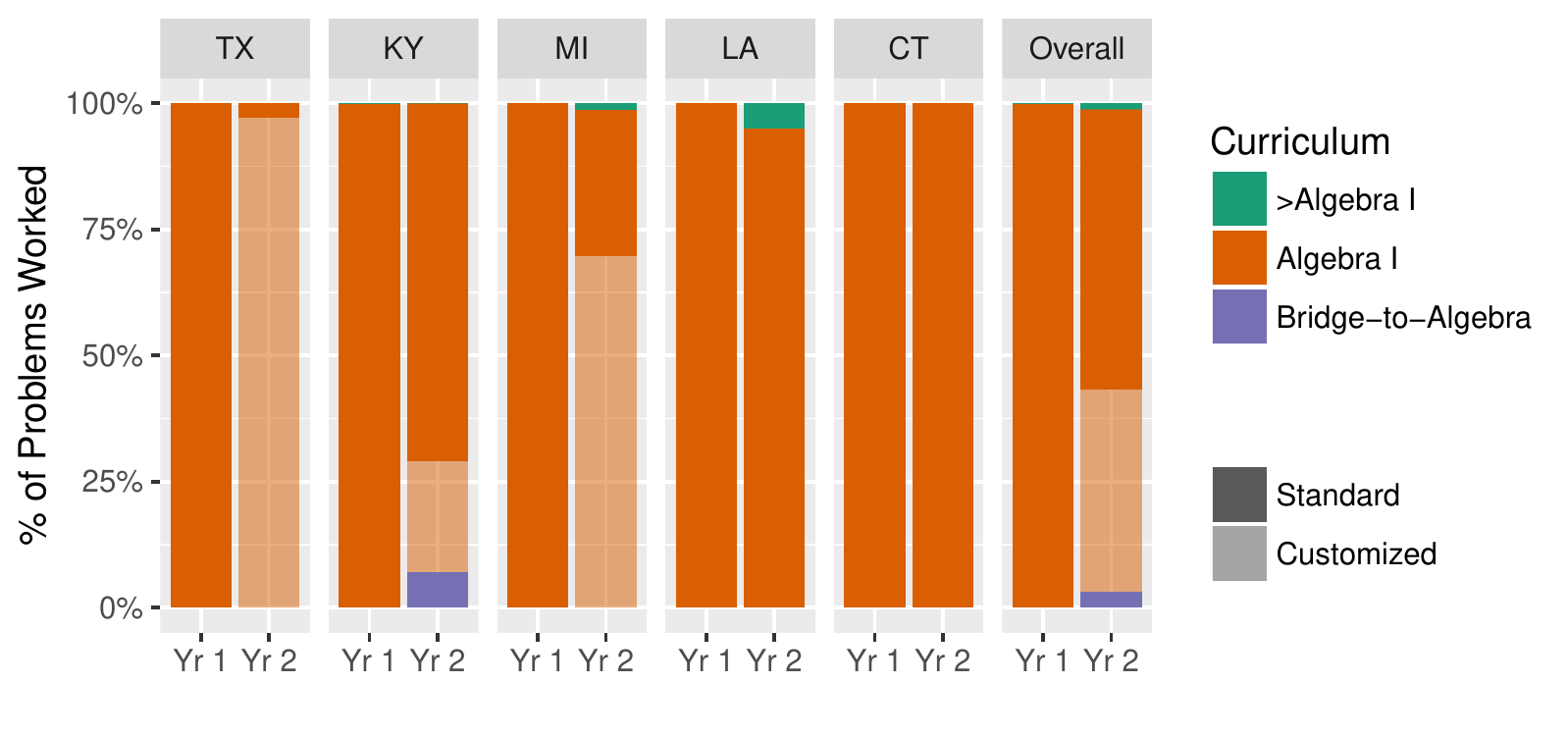}

\caption{Percentage of worked problems coming from various courses
  (denoted by color, with Algebra II and Geometry bundled as ``$>$Algebra I''), from
  standard and customized variants, denoted by shading.}
\label{fig:curricula}
\end{figure}

Students' automatic progress through the Cognitive Tutor (CT) software
is normally governed by
the sequences of sections and units embedded in the software.
Without external meddling, the curriculum a student works on
determines the sequence, and thus what section he or she will be directed towards next after
mastering (or exhausting the problems) from a previous section.
In the CTAI effectiveness trial, the most common curriculum was,
naturally, Algebra I.
This came with three closely related variants, due to new software releases.
Students requiring more remediation were able to work on a less
advanced curriculum, called ``Bridge to Algebra,'' and more advanced
students could work on Algebra II or Geometry.

In the second year of the study, some high schools, primarily in
Texas, Michigan, and Kentucky, requested customized variants of the
curricula.
This was typically due to state standards, testing schedules, or local
scope and sequence guidelines.
These ``customized curricula'' altered the order of some sections and
units, and were usually particular to schools.

Figure \ref{fig:curricula} shows the percentage of worked problems
from each curriculum, from standard and customized varieties, by state
and year.
First, note that the vast majority of worked problems were from the
Algebra I sequence.
A small but notable number of less advanced problems were worked in
Kentucky in year 2, and some more advanced problems were worked in
Michigan and Louisiana.
Secondly, note the rise in ``customized curricula'' in year 2 in
Texas, Kentucky, and Michigan, the three states with the most students
in our dataset.
In particular, Texas shifted almost entirely to customized curricula
from years 1 to 2.

Throughout the school year, teachers could have a class of students
working on multiple curricula either sequentially, where the students
changed curricula in lock step, or simultaneously, where students
worked on different curricula at the same time. As an example, two
teachers located in Kentucky had their students working on Algebra I
throughout most of the year and then reassigned them to Algebra II in
the last month of school. In contrast, a different teacher in Kentucky had students variously
enrolled in three different curricula throughout the entire year
(Bridge-to-Algebra, Algebra I, and a customized Geometry curriculum),
while a year 2 teacher in Michigan enrolled students in
three curricula sequentially throughout the year: Algebra I
until November, followed by a customized curriculum until February, and
ending with a different customized curriculum until June. While there are numerous instances of these
uses of multiple curricula in year 2, there are also many occurrences
of teachers who had their students enrolled in the standard Algebra I
throughout the entire year, including all Connecticut teachers.
There were also teachers, mostly in Texas, who used customized
curricula exclusively throughout the
second year.

\section{Student Usage Across States and Years}\label{sec:usage}

% latex table generated in R 3.3.1 by xtable 1.8-2 package
% Mon Nov  6 13:58:09 2017
\begin{table}[ht]
\centering
\begin{tabular}{rrrrr}
  \hline
 & Hours & Problems & Sections & Units \\
  \hline
Year 1 & 33.47 & 309 & 43.0 & 9 \\
  Year 2 & 23.83 & 228 & 36.5 & 10 \\
   \hline
\end{tabular}
\caption{Median numbers of hours, problems, sections, and units worked by each student in the dataset in the two years of the study. Students with no usage data were excluded.}
\label{tab:medUsage}
\end{table}

Table \ref{tab:medUsage} shows the median numbers of hours,
problems, sections, and units worked on by each student in the dataset
in the two years of the study.
Apparently, usage decreased markedly in the second year: the median
of hours worked decreased by
10, the median number of problems decreased by
81, and
the median number of sections decreased by
6.5 from
years 1 to 2.
Yet, as discussed below, the median number of units worked increased
by 1.

\begin{figure}
\centering

\includegraphics[width=\maxwidth]{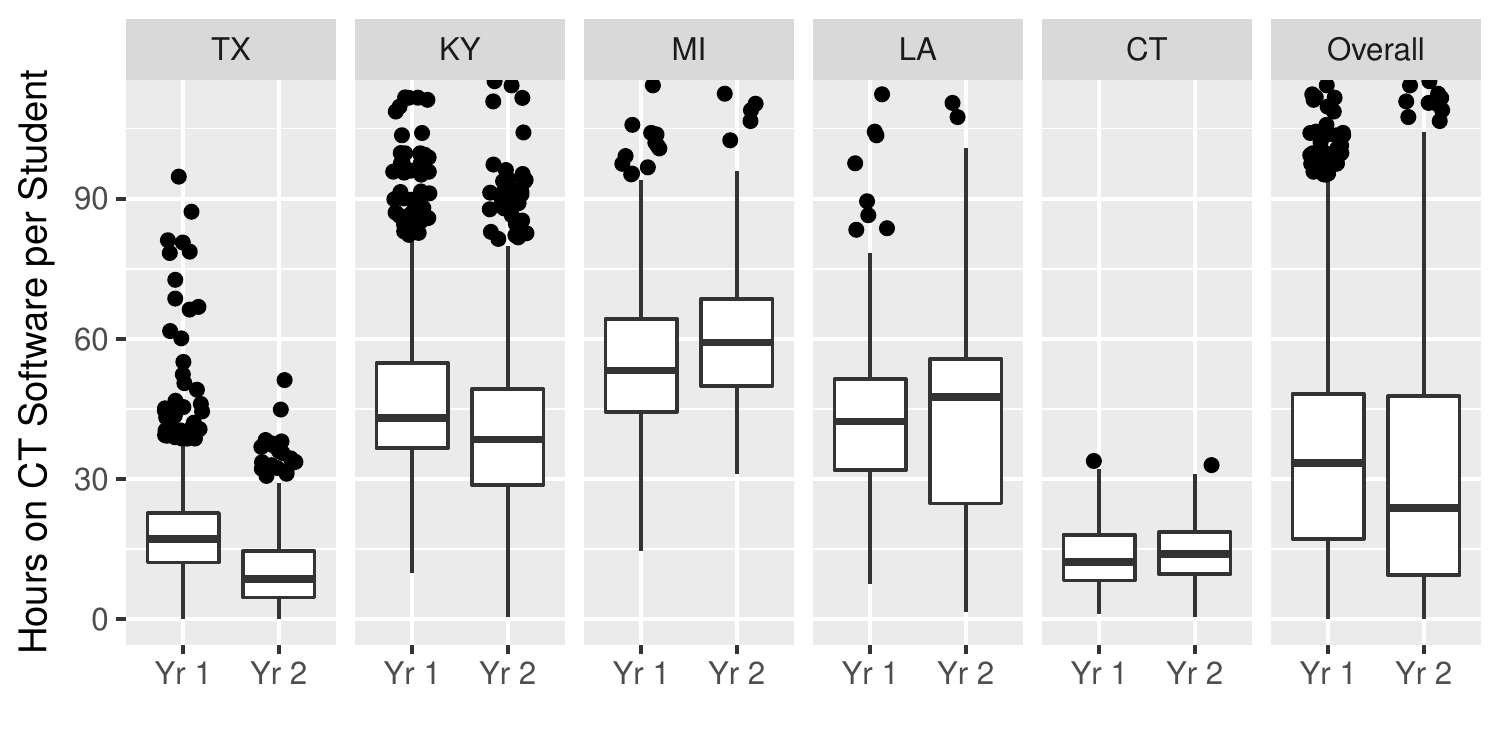}

\caption{Boxplots of hours each student spent on Cognitive Tutor
  software over by year and state. Students  with no timestamp data
  ($n=$193), with anomalous negative time
  ($n=$3) or with more than 110 hours ($n=$91)
  were excluded.}
\label{fig:timeByStud}
\end{figure}

Figure \ref{fig:timeByStud} shows that the number of hours students
spent working on the CT software in some more detail, via
state-by-year boxplots.
Analogous figures for the numbers of problems and sections students
worked,
showed similar patterns.
Usage time varied substantially between students and across states and
years.
Students in Texas,
Connecticut, and New Jersey worked far fewer hours than students in
Kentucky, Louisiana, and Michigan.
Not every state reduced its usage from years 1 to 2---while students
in Texas, Kentucky and New Jersey used the software less in the second
year than in the first, students in Michigan, Louisiana, and
Connecticut increased their usage.

Overall, usage varied a bit more in year 2 than in year
1---the median absolute deviation of time spent was 15.9 hours in the first year, compared to 16.9 in  the second year.
The increase in variation seems to be driven both by increasing
between-state variation, and a between-student increase in Louisiana.
One intriguing possibility is that the amount of CT usage may have been better tailored to
teachers and students in the second year than in the first. Perhaps
usage increased for
students who stood to gain more from the software and decreased for
students who stood to gain less.

\begin{figure}
\centering

\includegraphics[width=\maxwidth]{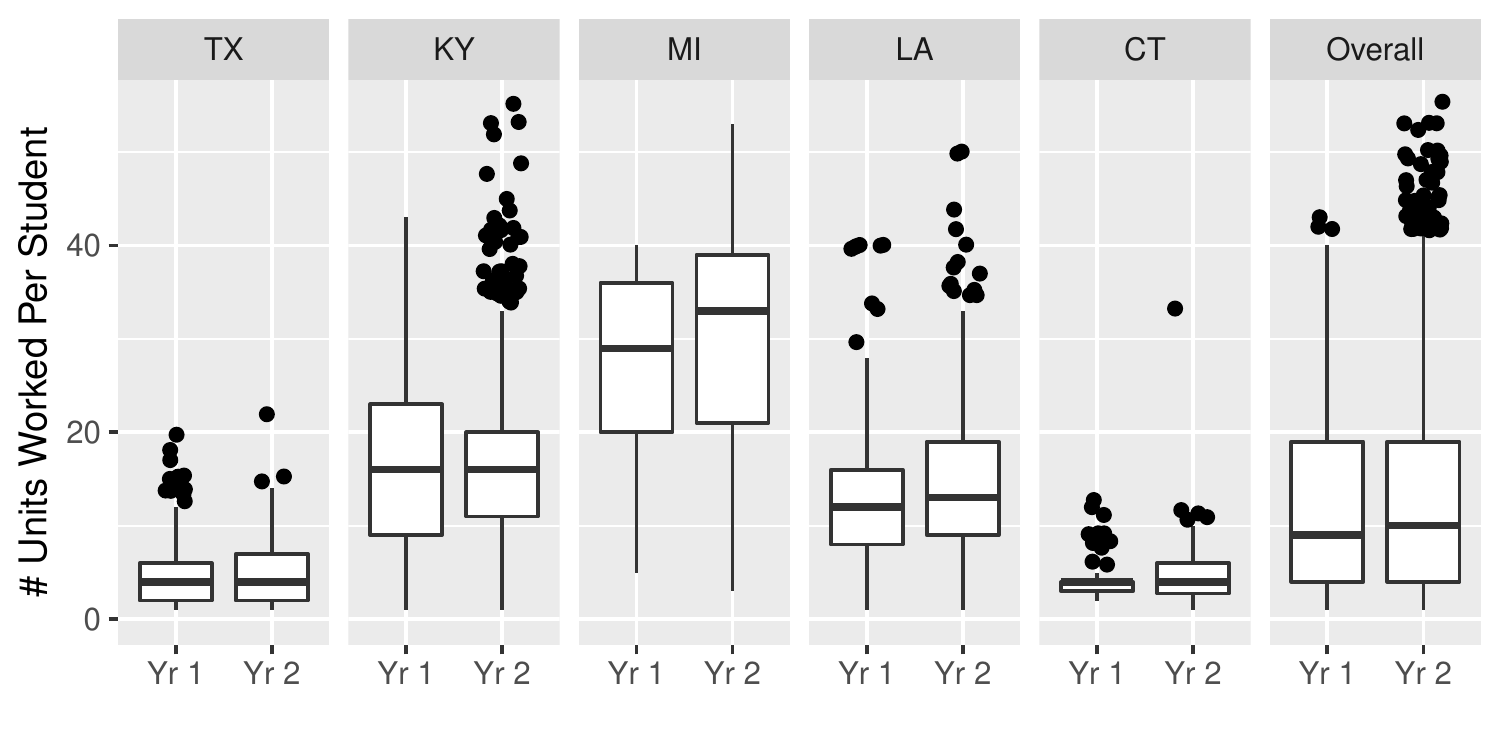}

\caption{Boxplots of the number of units of Cognitive Tutor
  software each student worked, by year and state. Students
  working more than 55 units (
  7 of
  4460) and students
  with no usage data (189)
  were excluded.}
\label{fig:unitsByStud}
\end{figure}

In contrast to the decreasing numbers of hours, problems, and sections students
worked in year 2, Table \ref{tab:medUsage} shows that the median number of units
students worked increased by
1
in year 2.
This suggests students in year 2 were exposed, on average, to a
slightly wider range of topics.
Figure \ref{fig:unitsByStud} shows boxplots of the numbers of units
worked by state and year.
The geographic variation in units worked mirrors the pattern in Figure
\ref{fig:timeByStud}, with more usage in Kentucky, Michigan, and
Louisiana but less in Texas, Connecticut, and New Jersey.
However, in every state the median year 2 student worked at least as
many different units as the median year 1 student.
Variation in the number of units worked also increased slightly
from years 1 to 2---the interquartile range (IQR) increased in every state
except for Kentucky, where a decrease in IQR was accompanied by an
increase in the number of outliers.

All in all, students used CT software less in year 2 than in year
1.
On the other hand, students in the second year tended to see a
slightly wider range of topics, and varied somewhat more in their usage.

\section{Working Units in Order---Or Not}\label{sec:order}

Overall, students used CT less in the second year than in the
first.
How was this difference distributed across CTAI units?

\begin{figure}
  \centering

\includegraphics[width=\maxwidth]{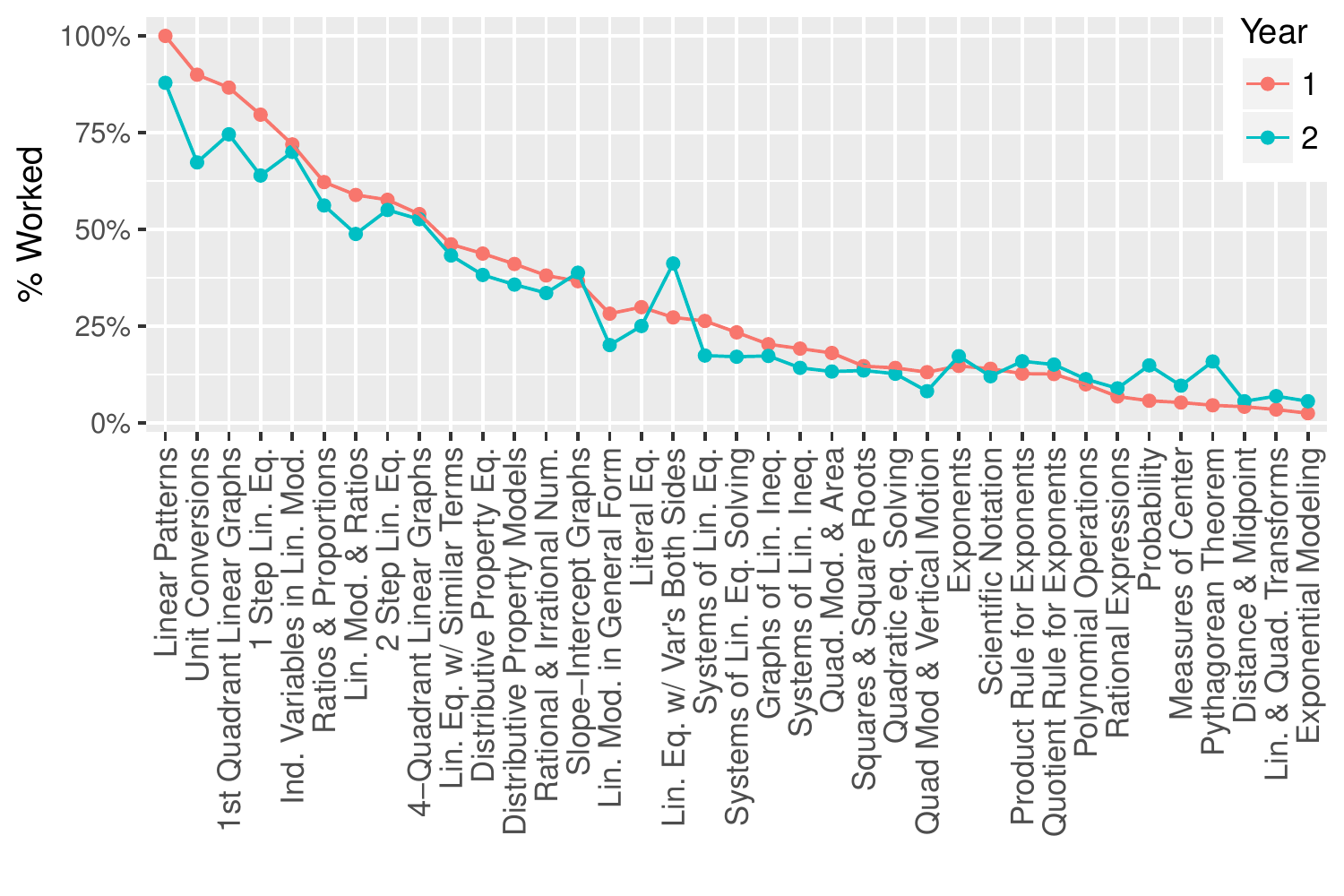}

\caption{The percentages of students with usage data who worked at
  least one problem from each unit in the Algebra I curriculum. The
  units are arranged in order for the standard curriculum. In the 2008
version of the software, the ``Unit Conversions'' unit was broken up
into two smaller units; for the sake of between-year comparisons, we
re-combined them.}
\label{fig:unitsWorked}
\end{figure}

Figure \ref{fig:unitsWorked} shows the units of algebra along the
horizontal axis, according to their order in the standard CTAI
curriculum.
The vertical axis shows the percentage of students with usage data who
worked each unit.

In year 1, the curve is almost monotonically decreasing, as one would
expect if students adhered to the curriculum.
Students varied in the number of units they worked---with the variation due
to both student ability and the amount of time allocated to CTAI
within a classroom---but they mostly followed the standard curriculum.
Students who worked fewer units stopped earlier in the sequence, and those who worked more units progressed farther.
Hence, earlier sections were worked by higher proportions of students
than later units.

In contrast, in year 2 students were much more likely to depart from the
standard unit order.
For instance, Figure \ref{fig:unitsWorked} suggests that some students
skipped ``Unit Conversions'' to work on  ``1st
Quadrant Linear Graphs'' or skipped ``1 step Linear Equations'' to work on ``Independent Variables in Linear Models''
In both these cases, the subsequent unit was worked on by a greater
proportion of students than the immediately prior unit.

Most strikingly, ``Linear Equations with Variables on Both Sides'' was worked by a
greater proportion of students in year 2 than in year 1, and by a
greater proportion of students than any of the previous six sections.
Presumably teachers and administrators wanted students to focus on
that unit, perhaps because they found it to be particularly effective,
because students tend to struggle with its main topic, or because its
topic may figure prominently in an upcoming standardized test.

\begin{figure}
  \centering

\includegraphics[width=\maxwidth]{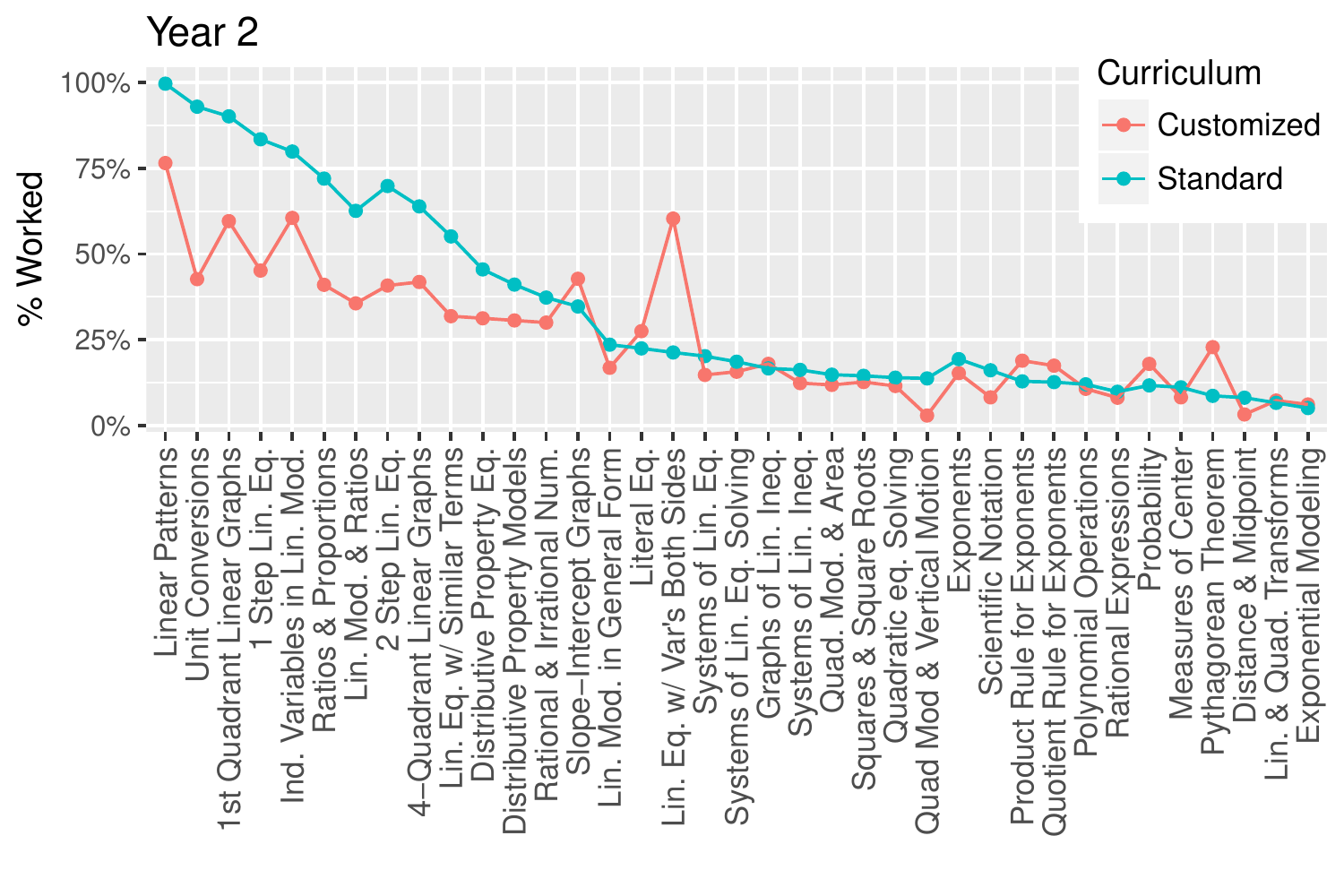}

\caption{The percentages of year-2 students with usage data who worked at
  least one problem from each unit in the Algebra I curriculum. The
  units are arranged in order for the standard curriculum. Students
  are divided between those attending schools using primarily a customized
  curriculum and those using primarily the standard Algebra I curriculum.}
\label{fig:unitsWorkedCust}
\end{figure}

Most of the variation in unit order was driven by the rise, in year 2,
of customized curricula.
Figure \ref{fig:unitsWorkedCust} divides year-2 students into those
attending schools using primarily a customized curriculum, and those
attending schools using primarily a standardized
curriculum.\footnote{At least
  90\% of
  problems worked by students at ``Customized'' schools were from a customized curriculum,
  and at most
  1\% of
  problems at ``Standard'' schools were from a customized curriculum.}
Students using a standardized curriculum followed the standard
sequence---more or less---while students using customized curricula
did not.
That said, there were some order violations in the standard group:
specifically, more students worked problems from units ``2-step Linear
Equations'' and ``Exponents'' than worked the preceding sections;
this suggests that some teachers used the reassignment tool to
prioritize particular topics.
Of course, teacher reassignment may have occurred in schools with customized
curricula as well---a possibility we will discuss in the next
section.

\begin{figure}
  \centering

\includegraphics[width=\maxwidth]{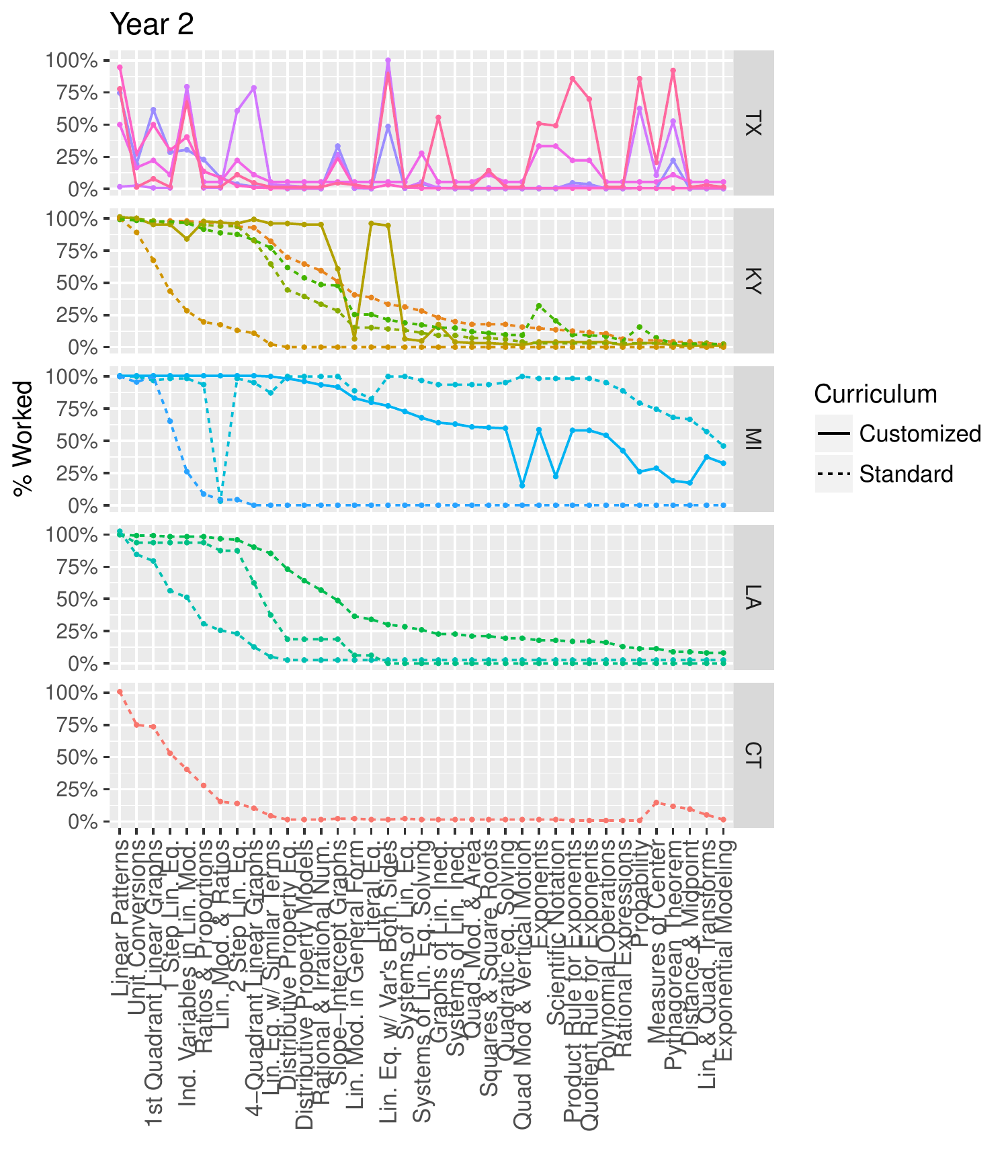}

\caption{The percentages of year-2 students with usage data in each
  school who worked at least one problem from each unit in the Algebra I curriculum. The
  units are arranged in order for the standard curriculum. Schools are
  classified as either using primarily customized curricula (solid
  line) or using primarily the standard Algebra I curriculum (dotted).}
\label{fig:unitsBySchool}
\end{figure}

Figure \ref{fig:unitsBySchool} further decomposes the year-2 results
by school and state, showing a large amount of variation between
states, as well as variation between schools within states.
In Texas, every school used customized curricula, most of which seem to
prioritize some of the same units, for instance, ``Linear Patterns,''
``Independent Variables in Linear Models,'' and
``Linear Equations with Variables on Both Sides''
On the other hand, there was also variance between schools.
For instance, one school prioritized units ``2 Step Linear Equations''
and ``4-Quadrant Linear Graphs'' while nearly eliminating ``Linear Patterns.''

Between-school variation is evident in the other states, as well.
In four of the five Kentucky schools, nearly every student worked on
the first nine units; in the one Kentucky school that used a
customized curriculum, nearly every student worked on the first 13
units, omitted the 15th (``Lin. Mod. in General Form''), and worked on
the 16th and 17th (``Literal Equations'' and ``Linear Equations with Variables on Both Sides'').
In the remaining school, nearly every student worked on the first
section, but usage decreased rapidly from there.
In one Michigan school which used the standard curriculum,
no students seem to have worked on the ``Linear Models \& Ratios'' section.

If unit order and topic scaffolding are important to CT's mastery
learning mechanism, the wide variation in students' realized curricula
would seem to pose a problem.
The fact that the prescribed order was followed less in the second
year of the study, when CTAI was effective, than in the first year,
when it wasn't, suggests that the standard curriculum may play a
smaller role than one might otherwise imagine.

\section{Mastering the Material---Or Not}\label{sec:mastery}
The central idea behind mastery learning is that students progress
through the curriculum as they master skills.
In the context of CT, skills are clustered within sections, which are
in turn clustered within units.
Students progress from the current section to the next section after
mastering all of the current section's skills.
Ideally, students would master all of the skills in all of the
sections they work.

By default, the software operates by automatically moving students
from section to section based on the sequence of topics defined by the
curriculum they were currently enrolled in. In this
software-controlled sequencing, students ideally spend the time
necessary to learn the material of a section, are judged by the
software to have mastered the material, and then ``graduate'' to the
next section.
However, the software will also ``promote'' a student to the next
section if the student exhausts a section's material without mastering
its skills.
Additionally, teachers are able to modify a student's path within the curriculum.
They can ``reassign'' students from their current sections to other
sections earlier or later in the intended sequence, including sections
they worked on previously.
Finally, if the semester ends, or a student stops using CT for some
other reason, while in the middle of working through a
section, the section is designated ``final.''
All in all, each CT section a student encounters ends in one of four possible ways:
mastery, promotion, reassignment, or as the student's final section.

\begin{figure}
  \centering

\includegraphics[width=\maxwidth]{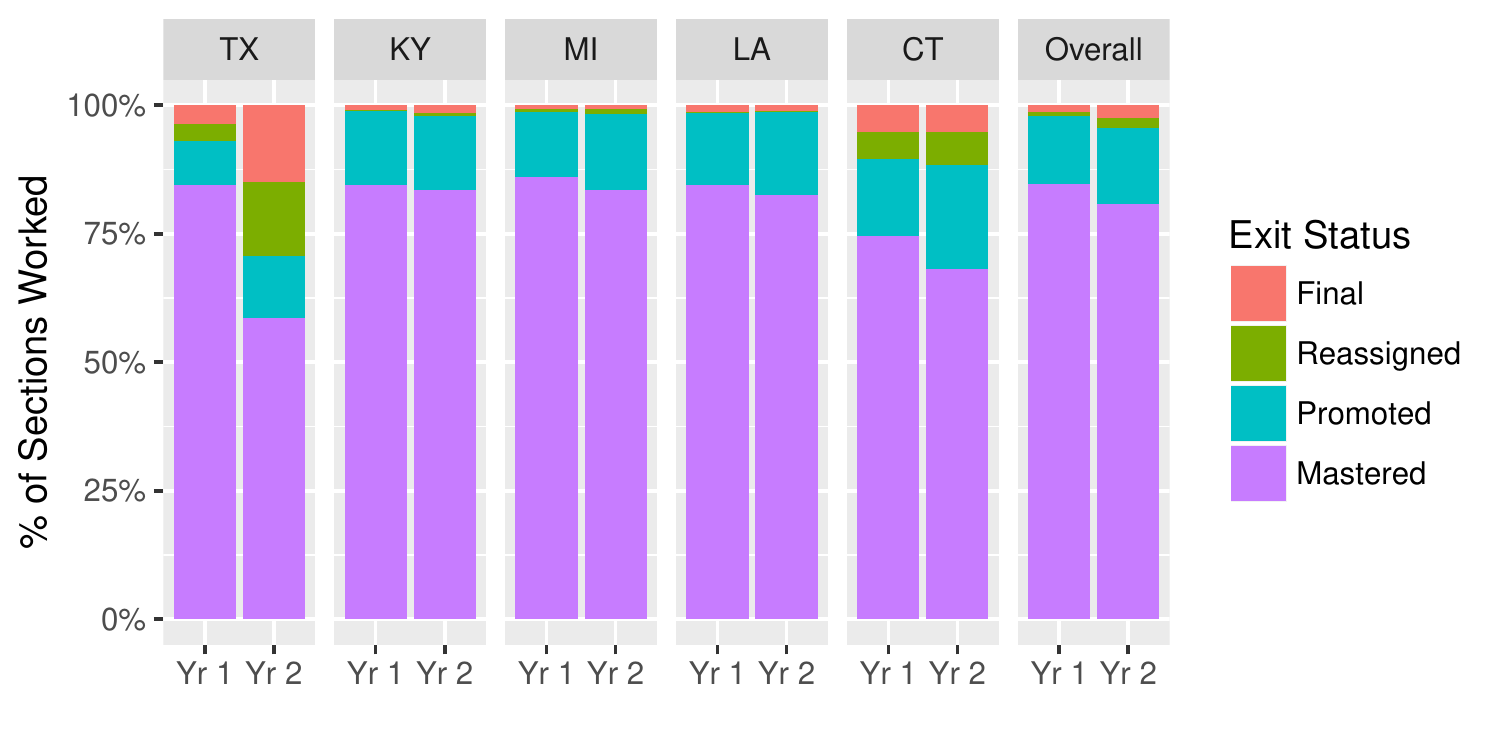}

\caption{The distributions of outcomes of worked sections, by state and across the
  entire sample, in the two study years.}
\label{fig:overallStatus}
\end{figure}

\begin{table}
  \centering
 \begin{tabular}{rllllll}%|llllll}

&   \multicolumn{6}{c}{Year 1}\\%&\multicolumn{6}{c}{Year 2}\\
&TX&KY&MI&LA&CT&Overall\\Final&699&625&242&129&91&1796\\Reassigned&600&140&220&5&92&1082\\Promoted&1619&9628&4582&1459&261&17561\\Mastered&15815&56199&31200&8648&1303&113374\\Total&18733&66592&36244&10241&1747&133813\\&\multicolumn{6}{c}{Year 2}\\
Final&1252&660&236&137&114&2412\\Reassigned&1215&253&275&9&144&2042\\Promoted&1000&6296&4717&1998&450&14477\\Mastered&4929&36676&26509&10108&1513&79874\\Total&8396&43885&31737&12252&2221&98805\\

\end{tabular}
\caption{Numbers of worked sections that ended in each of the
  four possible outcomes, across states and study years.}
\label{tab:overallStatus}
\end{table}

Figure \ref{fig:overallStatus} and Table \ref{tab:overallStatus} show
the proportions of worked sections in each state and study year that
ended with mastery, promotion, or reassignment, or as the
student's final section.
In the first year, about
85\% of
worked sections are mastered, except in Connecticut.
Other than in Texas, about
13--15\%
of sections end in promotion.
About 3\% of sections in Texas
and 5\%
in Connecticut end in reassignment, which is even rarer in the other states.

With the exception of Texas, sections tended to be completed similarly
in both years.
In Texas, however, the percentage of sections ending in reassignment increased by a factor
of about about
5,
to about 14\%.
The proportion of Texas sections labeled ``Final'' increased as
well---the expected result of decreasing the overall number of worked
sections and holding fixed the likelihood of ending
usage while in the middle of a section.

Across states, sections ended in reassignment at a rate of about
1\% in year 1 and
2\% in year 2.

\subsection{Section Mastery and Curriculum}
\begin{figure}
  \centering

\includegraphics[width=\maxwidth]{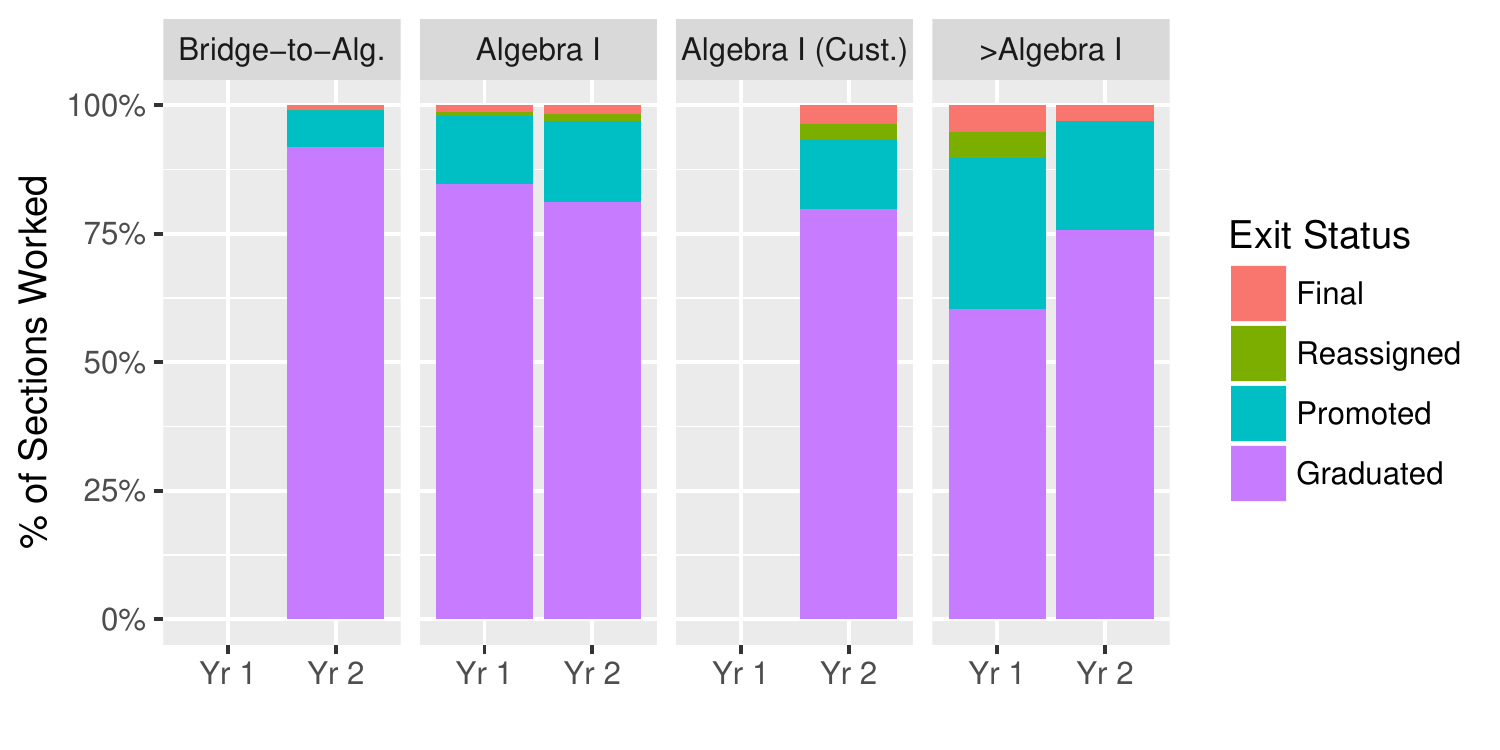}

\caption{The distributions of outcomes of worked sections, by
  curriculum, in the two study years. (There were no Bridge-to-Algebra
sections in customized curricula in our dataset.)}
\label{fig:statusCur}
\end{figure}

A well-designed curriculum, can, in theory, play an important role in
students' attainment of mastery.
Students who work on appropriate problems that build on their current
set of skills should be more likely to master new skills than students
working on problems above their level.
What role did variations in the CT curriculum play in mastery during the effectiveness trial?

Figure \ref{fig:statusCur} shows the proportions of worked sections
that were mastered or ended in promotion, reassignment, or finality, in
standard and customized versions of each CT curriculum.
Mastery proportions do, indeed, depend on curriculum.
Specifically, students mastered sections from more advanced curricula
less frequently.
Sections from the most basic curriculum, Bridge to Algebra, were
mastered
92\%
of the time;
those from Algebra I were mastered
83\%
of the time, and those from more advanced curricula were mastered
at a rate of
75\%.
This is unsurprising, since more advanced curricula may be expected
to be more challenging.
However, it may suggest that some students studying advanced topics
would fare better in more standard curricula.

Algebra I sections from customized curricula tended to end in
reassignment more often than sections from the standard Algebra I
curriculum (3\% vs.
1.4\%, in year 2).
This may indicate an overall skepticism towards the Carnegie Learning standards
among certain schools and teachers, manifested in both adoption of
alternative curricula and reassignment.%\marginpar{does that make more sense?}

\section{Digging Deeper into Section Reassignment}\label{sec:cp}
The proportion of worked sections in our dataset ending in
reassignment was small.
Nevertheless, since reassignment represents the only mechanism by
which individual teachers can affect their students' progress through
the Cognitive Tutor, exploring patterns of reassignment can provide
insight into how CT was used.

\subsection{How Do Reassignment Patterns Vary?}

\begin{figure}
  \centering

\includegraphics[width=\maxwidth]{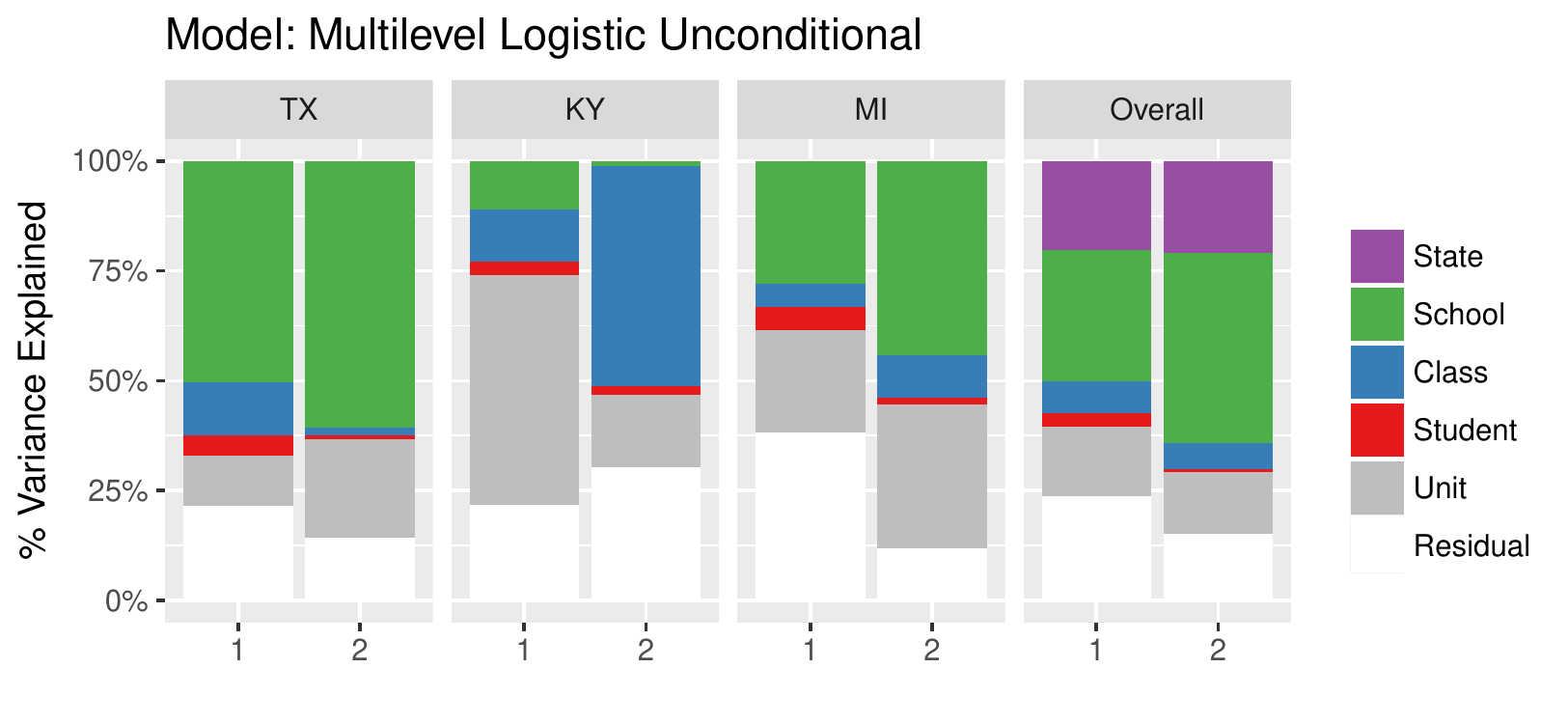}

\caption{Results from a set of eight multilevel logistic regressions
  predicting section reassignment. For each year, in the entire sample (``Overall'')
  and in the three states with the largest numbers of reassignments (Texas,
  Kentucky, and Michigan), we regressed a binary variable indicating
  whether a section ended in reassignment on random intercepts for
  school, class, student, and unit, and in the overall case, for state
  as well, and recorded their variance. The residual variance was set
  as the variance of the standard logistic distribution,
  $\pi/3$. These bar charts give the proportion of the total variance
  attributable to each random effect.}
\label{fig:vc}
\end{figure}
Teachers alone control reassignment.
Nevertheless, the factors influencing student reassignment vary at a
number of levels.
For instance, state and district standards may prod teachers into
reassigning students to particular units.
Some principals may encourage teachers to adhere to the official
curriculum and avoid reassignment.
Some students may be more prone to reassignment than others.
Certain units in the CTAI curriculum may be harder than others,
causing students to tarry and teachers to reassign.
Finally, a host of other factors, at these levels and others, may spur reassignment.

To better understand the source of the variation in
reassignment---what drives some, but not other, sections worked by
students to end in
reassignment---we fit a set of multilevel models.
We fit separate models to data from each the three states with the highest numbers of reassignments, Texas,
Kentucky, and Michigan, and in the sample as a whole, in each of the
two study years, yielding a total of eight models.
Each model was a logistic regression: a binary indicator for section
reassignment was regressed on a random intercept for unit, as well as
nested random intercepts for student, classroom, and school.
Models fit to data from all six states included an additional random
intercept for state.

Logistic regression can be represented in terms of an underlying
latent variable $Z^*$: student $i$ working section $sec$ is reassigned
when $Z_{sec,i}^*>0$.
The model for $Z^*$ is:
\begin{equation*}
 Z^*_{sec,i}=\alpha_0+\beta_{u[sec]}+\gamma_i+\delta_{c[i]}+\epsilon_{s[i]}+e_{sec,i}
\end{equation*}
Where $\alpha_0$ is an overall intercept,
and $\beta_{u[sec]}$, $\gamma_i$, $\delta_{c[i]}$, and $\epsilon_{s[i]}$
are random intercepts for the unit in which $sec$ appears, for student
$i$, for $i$'s classroom, and for $i$'s school, respectively.
Again, the model fit to all six states includes an additional random
intercept for state.
The random intercepts are modeled as independent and normally
distributed, each with its own variance.
The regression error $e_{sec,i}$ is given the standard logistic
distribution, with ``residual'' variance $\pi/3$.
It is convenient to represent variance in reassignment probabilities
in terms of the variance of $Z^*$.

Figure \ref{fig:vc} gives the variance components estimated from these
logistic regressions: variances of the random intercept terms, as a
percentage of the total variance of $Z^*$.
Overall, in both years of the study, the largest determinant of
reassignment was school, accounting for
30\% of the variation in year 1,
and 43\% in year 2.
After school, state was the most important, accounting for
20\% and
21\% in the two years, and unit,
accounting for 16\% and 14\%.
Surprisingly, classroom and student-level factors only accounted for
7\% and
3\% in year 1, respectively,
and 6\% and
1\% in year 2.
The pattern was similar in Texas---where school accounted for over half
the variation in reassignment in both years---and in Michigan to a
lesser extent.\footnote{Percentages in state specific models, in which
  there is no between-state variance, cannot be
  directly compared to those from the overall model.}
In Kentucky, unit played the largest role
(52\%) in year 1, and classroom played
the largest role in year 2 (50\%).
Across states and years, student level factors never accounted for
more than 5\% of
the variation in reassignment.
Other than in Kentucky in year 2, classroom never accounted for more
than 12\%  of
the variation.

\textbf{Summary.} Although teachers control reassignment, their decisions appear to be largely
determined by broader policies, occurring at the state or school level.

\subsection{When are Students Reassigned?}

\begin{figure}
  \centering

\includegraphics[width=\maxwidth]{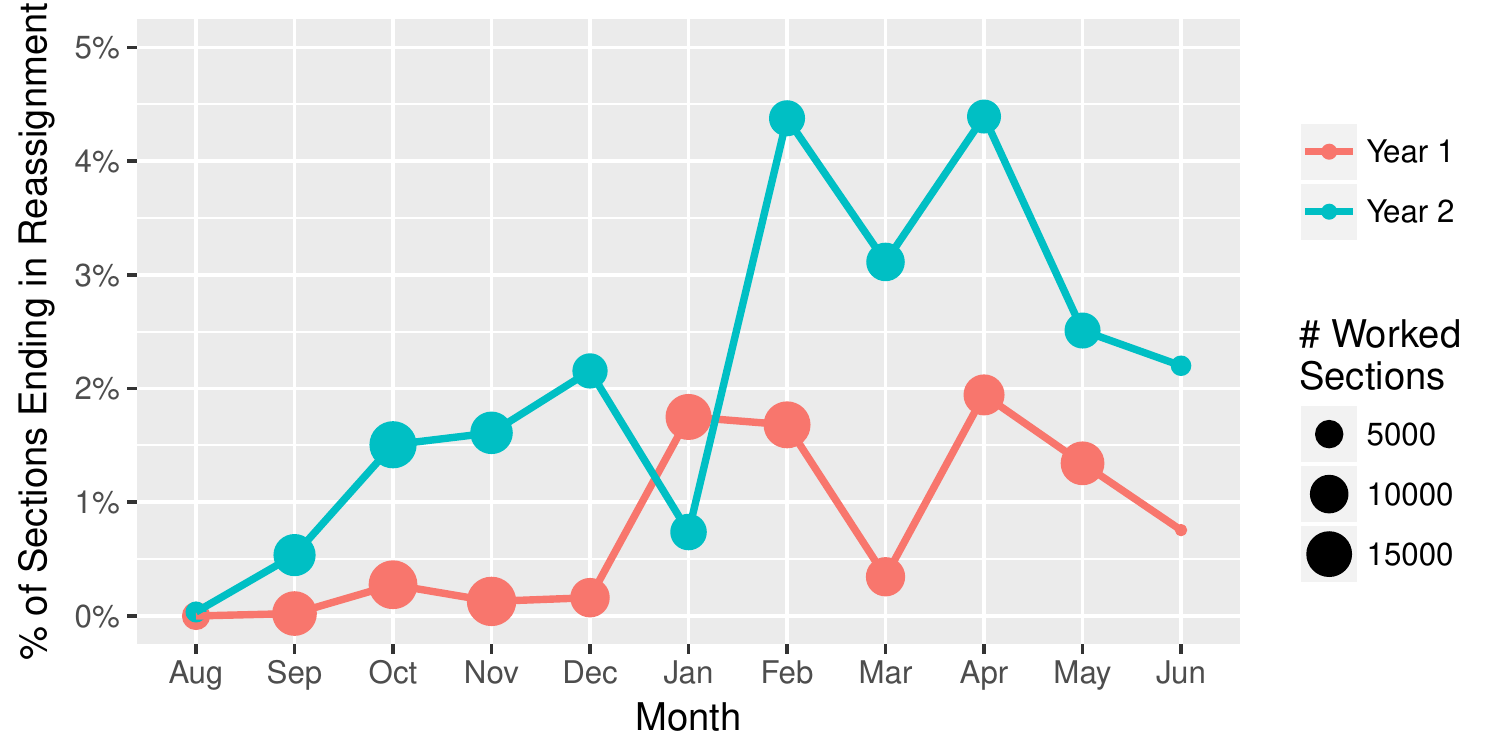}

\caption{The proportion of worked sections ending in reassignment, by
  month and year.}
\label{fig:byMonth}
\end{figure}

The timing of reassignments can also provide a window into what drives
teachers' decisions to reassign students.
Figure \ref{fig:byMonth} shows the proportion of worked sections in
each month that end in reassignment.
In both years, reassignments were much more common in the second half
of the school year than in the first.
This may be the result of teachers learning how to use the software as
the year progresses, or responding the pressure of upcoming
standardized tests by accelerating students' progress and reassigning
students to relevant sections.

As we've seen, reassignment was more common in year 2 than in year 1.
In fact, reassignment increases fairly steadily over the entire length
of the study.
Through December of the first year, reassignment was rare. From
January through May of year 1, between one and two percent of sections
ended in reassignment.
Year 2 begin where year 1 left off, with one to two percent of
sections reassigned.
Finally, from February through May of the second year, the rate of
reassignment increased again.

\begin{table}
  \centering
  \begin{tabular}{rll|ll}
    &\multicolumn{2}{c}{Year 1}&\multicolumn{2}{c}{Year 2}\\
    &Aug--Dec&Jan--Jun&Aug--Dec&Jan--Jun\\
% latex table generated in R 3.3.1 by xtable 1.8-2 package
% Mon Nov  6 14:19:52 2017
 TX &  18 & 582 & 281 & 932 \\
  KY &   9 & 131 &  52 & 201 \\
  MI &  47 & 173 & 245 &  30 \\
  LA &   0 &   5 &   9 &   0 \\
  CT &  14 &  78 &  84 &  60 \\
  NJ &   3 &  22 &  17 & 129 \\
   \hline
Overall &  91 & 991 & 688 & 1352 \\

\hline
\end{tabular}
\caption{The number of reassignments in each state and overall in the
  first and second halves of the year}
\label{tab:byMonth}
\end{table}

\begin{figure}
  \centering

\includegraphics[width=\maxwidth]{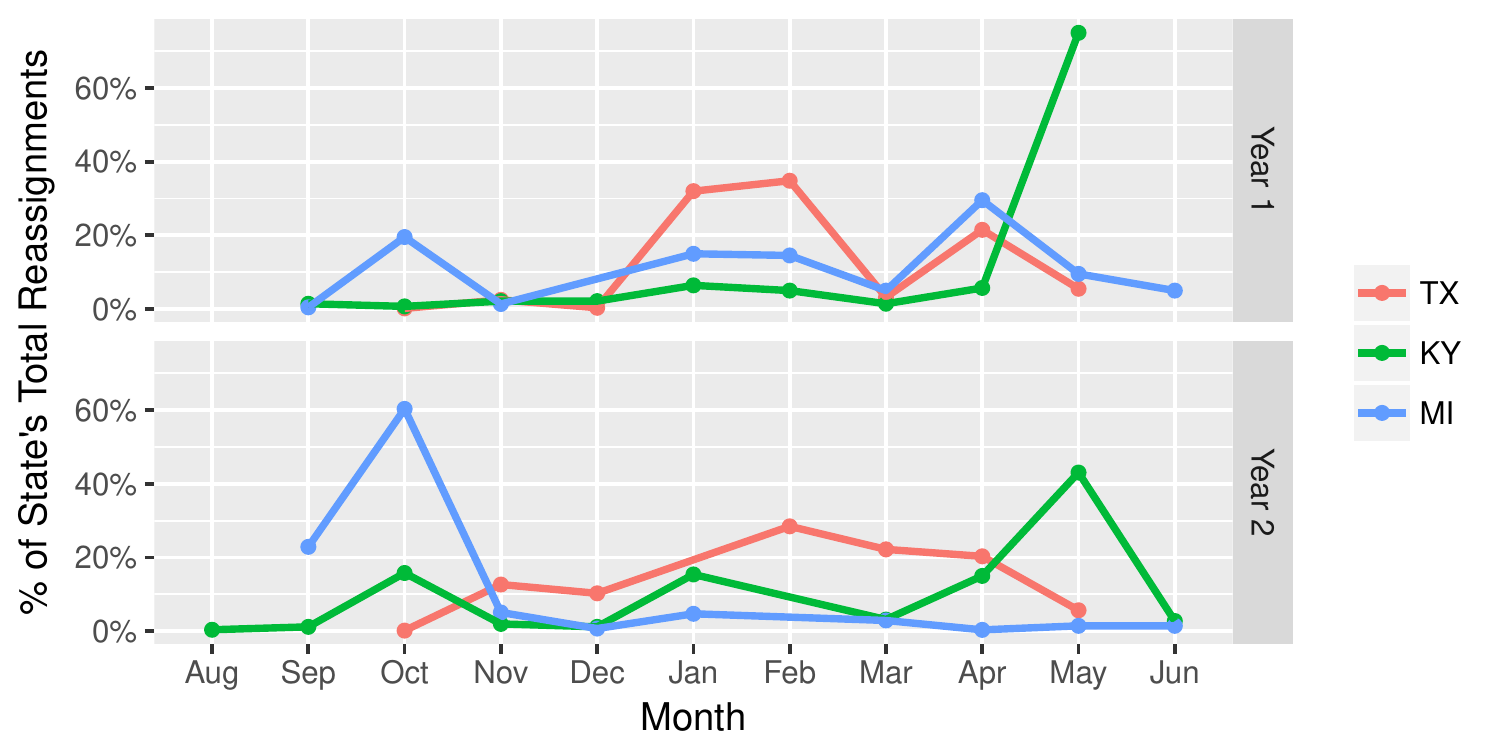}

\caption{The proportion of a state's reassignments that occurred in
  each month of the year.}
\label{fig:byMonthState}
\end{figure}

Figure \ref{fig:byMonthState} and Table \ref{tab:byMonth} decompose
these trends by state, Figure \ref{fig:byMonthState} includes just
the three states with the highest number of reassignments, Texas,
Kentucky, and Michigan. Table \ref{tab:byMonth} shows data for each of
the six states and overall.
For the two study years, Figure \ref{fig:byMonthState} shows the proportion of all of each
state's reassignments the occurred in each month of the school year.
(Note that while both Figure \ref{fig:byMonth} and \ref{fig:byMonthState}
show proportions for each month, the denominators are not the same:
Figure \ref{fig:byMonth} gives the proportion of each month's worked
sections that ended in reassignment, and Figure \ref{fig:byMonthState}
gives the proportion of all of the state's reassignments over the
course of the year, that occurred in each month.)

Figure \ref{fig:byMonthState} reveals patterns that are not apparent
in Figure 10.
For example, the bulk of Kentucky's reassignments took place in May.
A fifth of Michigan's reassignments in year 1 occurred in October;
for the rest of the year, Michigan roughly followed the same pattern as Texas.
In year 2, the vast majority of Michigan's reassignments also occurred near
the beginning of the year, in September and October.

\textbf{Summary.} In Texas and overall, reassignments were more common
in the second half of the school year than in the first half. In
Kentucky and Michigan, they were clustered in a few specific months.

\subsection{Does Reassignment Depend on Classmates?}

Student individuality and independence might be the most important
motivating factors behind mastery learning---each student learns at
his or her own pace, and struggles on a unique set of skills.
Students are supposed to move through the CT curriculum independently
of each other.
However, reassignments give teachers the ability to override this
feature, and coordinate students' progress.
Teachers might identify students who are behind their classmates and
reassign them to later sections.
They may also move an entire class together to a particular section or
unit of interest.
To what extent did these and similar considerations drive reassignment
in the CTAI study?

\begin{figure}
  \centering

\includegraphics[width=\maxwidth]{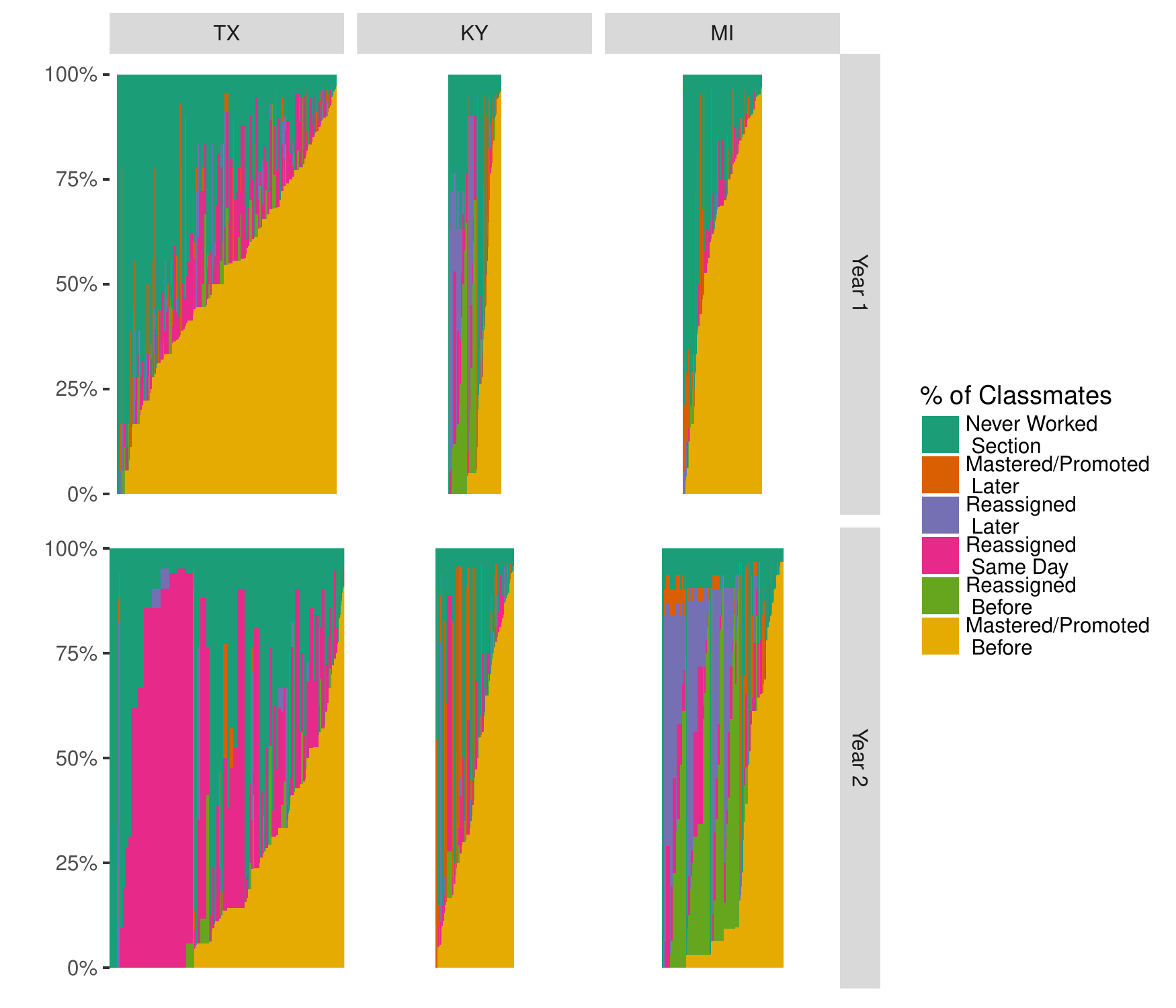}

\caption{Distributions of classmates' statuses at each reassignment in
  Texas, Kentucky, and Michigan. Each reassignment that took place in
  these three states is represented by a bar showing the proportions of
  the reassigned student's classmates who had exited the section on
  the same day or earlier via promotion or mastery, who had been
  reassigned from that section on an earlier date, on the same date,
  or on a later date, who mastered the section or were promoted on a
  later date, or who never worked the section at all. The bars are
  rank ordered according to those proportions, in the order listed.}
\label{fig:classmates}
\end{figure}

Figure \ref{fig:classmates} addresses this question by plotting a
students' classmates' statuses at the time he or she is reassigned.
For each reassignment in Texas, Kentucky, and Michigan, Figure \ref{fig:classmates}
plots a vertical bar colored to show the proportions of the reassigned
students' classmates (represented in the usage data) who  had exited the section on
the same day or earlier via promotion or mastery, who had been
reassigned from that section on an earlier date, on the same date,
or on a later date, who mastered the section or were promoted on a
later date, or who never worked the section at all.
The bars are rank ordered according to those same proportions: first by the
proportion of classmates who were promoted or mastered the section on
the same day as the reassignment in question, or earlier, next by the
proportion who were reassigned from the same section on an earlier
date, next by the proportion who were reassigned from the same section
on the same date, and finally by the proportion who were reassigned
from the same section on a later date.

Do teachers reassign students in order to help them catch up with classmates?
According to Figure \ref{fig:classmates}, that might be part of the
story, but isn't all of it.
Across years and states, in about 40\% of
reassignments, at least 75\% of the rest of the class had exited the
section through graduation or promotion on the same date or earlier.
The figure also reveals that the proportions of students who had graduated or been
promoted from the same section on the same day or earlier was smaller in year 2
than in year 1, especially in Texas and Michigan.

In Texas, year 2 saw a dramatic increase in the proportions of
students reassigned from the same section on the same day, suggesting
that some teachers may have been moving the class together through the
curriculum.
In Michigan in year 2, teachers reassigned almost all students who
worked certain sections.
That is, in 65\% of the instances in which year-two Michigan
students were reassigned from a section, at least 75\% of their
classmates were, at some point, reassigned from the same section, or
never worked it.
Across all three states and both years, it is exceedingly rare for
students to master or be promoted from a section after someone in
their class has been reassigned from the same section.

\textbf{Summary.} Three different patterns emerge from Figure
\ref{fig:classmates}: teachers reassigning students who have fallen
behind their classmates, teachers reassigning an entire class from the
same section on the same day, and teachers reassigning almost all
students who begin to work on particular sections.
Each of these patterns mostly takes place in different states and
years.

\subsection{Where To?}

Teachers who reassign students may simply move them to the next
section within the same unit.
Say that a teacher believes that a particular student had already mastered
the skills in one of the CTAI sections or is wheel-spinning---working
problems without learning---or a teacher dislikes one of the sections in a CT unit.
Still, the teacher wants the student to learn as much as possible from
the current unit.
Then moving the student to the next section within the same unit might
make sense.
Alternatively, teachers who believe that some of their students are
not progressing quickly enough may reassign them out of their current
units entirely, and into the next unit in the sequence---whatever that
may be.
Finally, a teacher who wanted his or her students to focus on a
particular topic might reassign them all to the appropriate unit
when the time is appropriate.

Which of these patterns is most prevalent?
More generally, when teachers reassign their students, where in the
curriculum do they send them?

\begin{figure}
  \centering
  \begin{subfigure}{5in}

\includegraphics[width=\maxwidth]{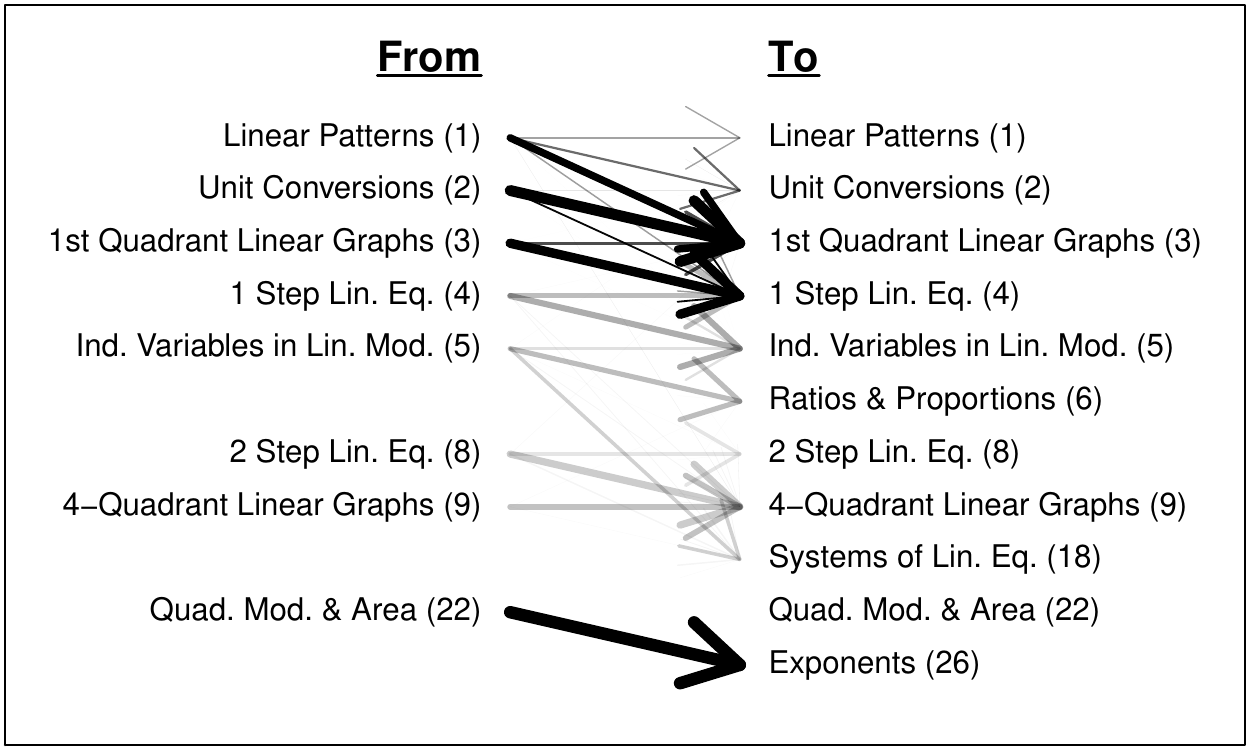}

\end{subfigure}
\begin{subfigure}{1in}
  \centering

\includegraphics[width=\maxwidth]{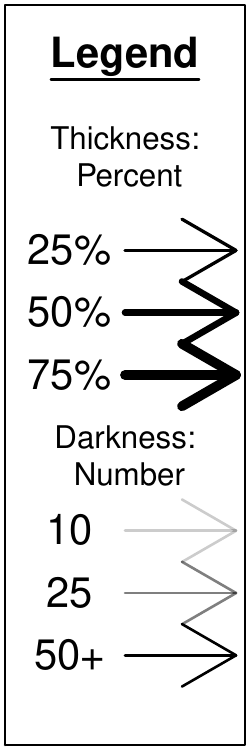}

\end{subfigure}
\caption{Reassignment transition plot for study year 1. The units on
  the left of the plot are the Algebra I units that ended in reassignment at
  least 30 times in year 1. The units on the right are those that were
  the destination of at least 30 reassignments. Also included on the
  left are the top two sending units for each of the units on the
  right, and also included on the right are all the units on the left,
  along with those units' top two receiving units. The units are numbered
  according their order in the standard Algebra I curriculum. There is an arrow from a
  unit on the left to a unit on the right if a student was assigned
  from the unit on the left to the one on the right. The thickness of
  the arrows is proportional to the percentage of all reassignments from
  the sending unit that ended in that receiving unit. The darkness of
  the arrows is proportional to the number of reassignments from the
  sending unit to the receiving unit.}
\label{fig:trans1}
\end{figure}

\begin{figure}
  \centering
  \begin{subfigure}{5in}

\includegraphics[width=\maxwidth]{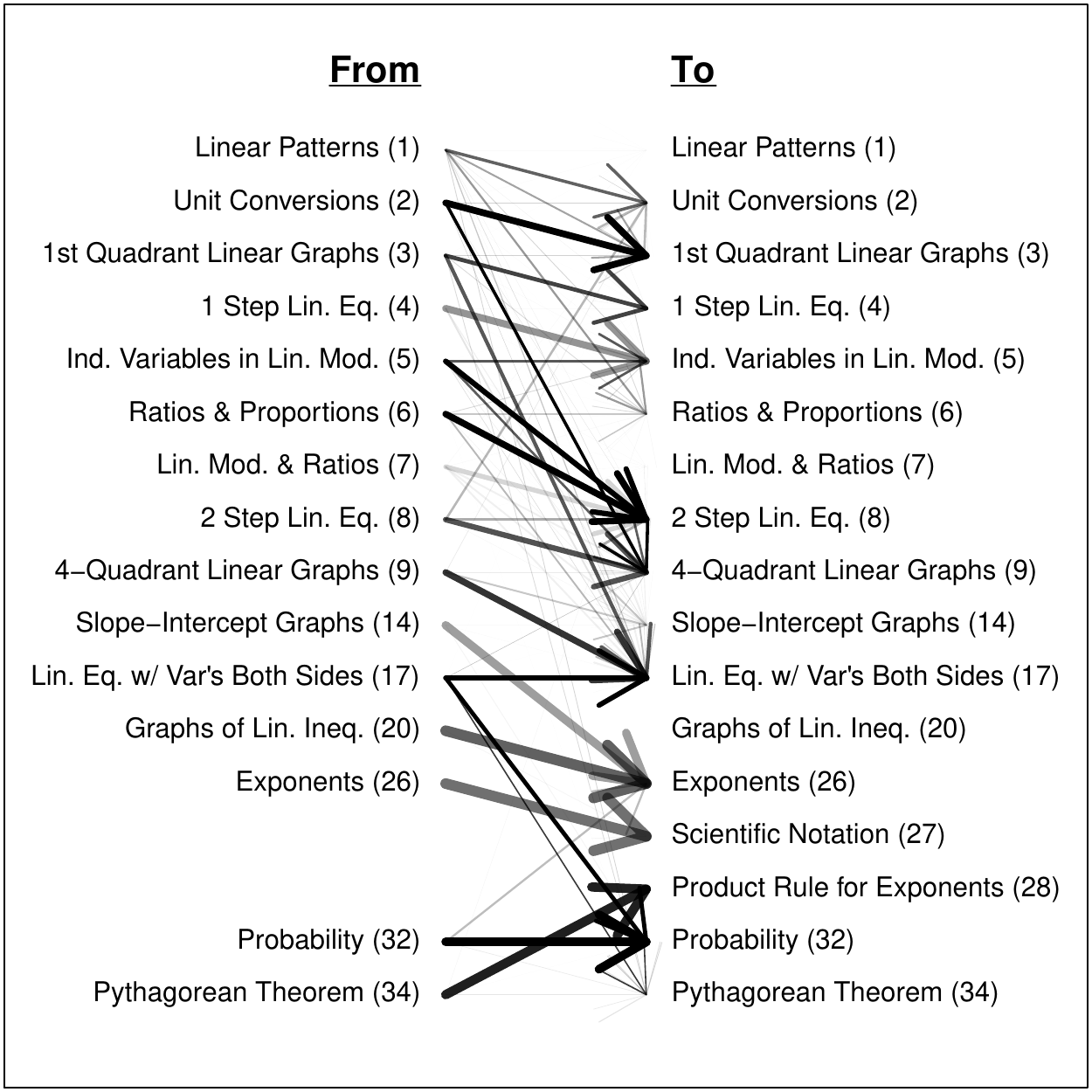}

\end{subfigure}
\begin{subfigure}{1in}

\includegraphics[width=\maxwidth]{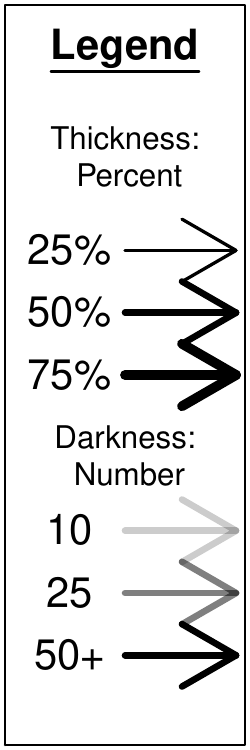}

\end{subfigure}
\caption{Reassignment transition plot for study year 2. See caption of
Figure \ref{fig:trans1} for details.}
\label{fig:trans2}
\end{figure}

To address these questions, we focus on the units in the standard
Algebra I sequence.
Figures \ref{fig:trans1} and \ref{fig:trans2} give transition plots
for reassignment in the two study years.
On the left of each figure, marked ``From,'' are the top ``sending''
units: the units that students were reassigned \emph{from} at least 30
times.
They are numbered according to the order they appear in the standard Algebra I curriculum.
On the right, under ``To,'' are the top ``receiving'' units: the units
that students were reassigned \emph{to} at least 30 times.
For completeness, each of the top two receivers for each sending unit,
and each of the top two senders for each receiving unit were also included.
Finally, all of the sending units listed on the left were also included in the
receiving column.
All in all, 75\% of the first-year reassignments and
75\% of the second-year reassignments are
captured in the figures.

The arrows in the plot represent reassignments.
An arrow from a sending unit to a receiving unit indicates
reassignment from the former to the latter.
The thickness of the arrows represents the proportion of reassignments
originating in the sending unit whose destination was the receiving unit.
The darkness of the arrows represents the number of such reassignments.
For instance, in Figure \ref{fig:trans1}, the arrow from
``4-Quadrant Linear Graphs'' on the left to ``4-Quadrant Linear
Graphs'' on the right is fairly thick, since
44\%
of the reassignments from ``4-Quadrant Linear Graphs'' end in the
same unit.
Yet it is also fairly faint, since it only represents
12
reassignments.

Inspecting the figures shows that the most common pattern is for
students to be reassigned to the next unit in the curriculum---this pattern comprises
52\% of the reassignments in year 1 and
24\% of those in
year 2.
On the other hand, it is relatively rare for students to be reassigned within the same unit
(these reassignments account for 15\% and 18\% in the two years).
It is also rare for students to be reassigned to units earlier in the curriculum (comprising
2\% and
8\%).

The transition plots also reveal some interesting cases worth highlighting.
In year 1 (Figure \ref{fig:trans1}), students reassigned from the first unit, ``Linear Patterns'', were primarily placed two units ahead, in ``1st Quadrant Linear Graphs,'' skipping ``Unit Conversions,'' perhaps suggesting a disinterest in ``Unit Conversions'' on the part of the reassigning teachers.
Similarly,
9
of the students
(27\%)
reassigned from the section ``Independent Variables in Linear Models'' were placed 13 units later in ``Systems of Linear Equations''
and all 62 of the students reassigned from ``Quadratic Models \& Area'' were placed in ``Exponents.''
This suggests that some teachers may have considered the units ``Systems of Linear Equations'' and ``Exponents'' to contain particularly important material.

Since the total number of reassignments was higher in year 2, the corresponding plot (Figure \ref{fig:trans2}) is larger and more complex.
It is also more common in year 2 for students to be reassigned to units other than the next unit in the sequences.
This may be partly due to the proliferation of customized curricula.
Two units, ``4-Quadrant Linear Graphs'' and ``Lin. Equations with
Variables on Both Sides,'' were common destinations from a wide
variety of earlier units, suggesting strong teacher interest in those units.
The majority (76\%)
of the students reassigned from ``Probability'' were placed in another section of the same unit, perhaps suggesting problems with some of that unit's sections; in fact, all of the students reassigned within the ``Probability'' unit were reassigned from one its first three sections (of seven).
Finally, 44
(81\%) of the reassignments from ``Pythagorean Theorem'' ended in the earlier ``Product Rule for Exponents'' unit.

\textbf{Summary.} The most prevalent pattern was for students to be reassigned
to the following unit, suggesting that teachers may be mostly
interested in advancing students who are behind.
On the other hand, a number of examples of other patterns---students moving within the same unit, or to units out of sequence---appear as well, suggesting that some teachers may be finely manipulating their students' curricula.

\section{Effects of Reassignment}\label{sec:effects}
The goal of CTAI is to help students learn Algebra, so
the most important questions about reassignment are about its effect on learning.
Although the data from this study came from a randomized trial, it was
CTAI as a whole that was randomized, not individual behaviors within CTAI.
Specifically, student reassignment was not randomized.
Therefore, precise estimates of causal effects of reassignment on
learning require strong untestable assumptions that are unlikely to be
true.
As in all observational studies, this includes the assumption that all
confounding variables---variables that predict both reassignment and
learning---have been measured well and modeled correctly.
Further complicating matters, although reassignment itself is a
well-defined process, in practice it can take many forms, as we have
endeavored to show.
There is no reason to expect the effects of reassignment to be
the same regardless of whether the teacher used it to help lagging
students catch up, to allocate time to important topics, or for some
other reason.

All that said, observational estimates of reassignment's average
causal effects can be valuable, if interpreted cautiously, for
instance by assessing their sensitivity to unmeasured confounding, as
we do below.
In the absence of evidence from randomized trials, observational
studies can help guide intuition, future research, and even---when
combined with other relevant information and theory---practice.

\begin{figure}
  \centering

\includegraphics[width=\maxwidth]{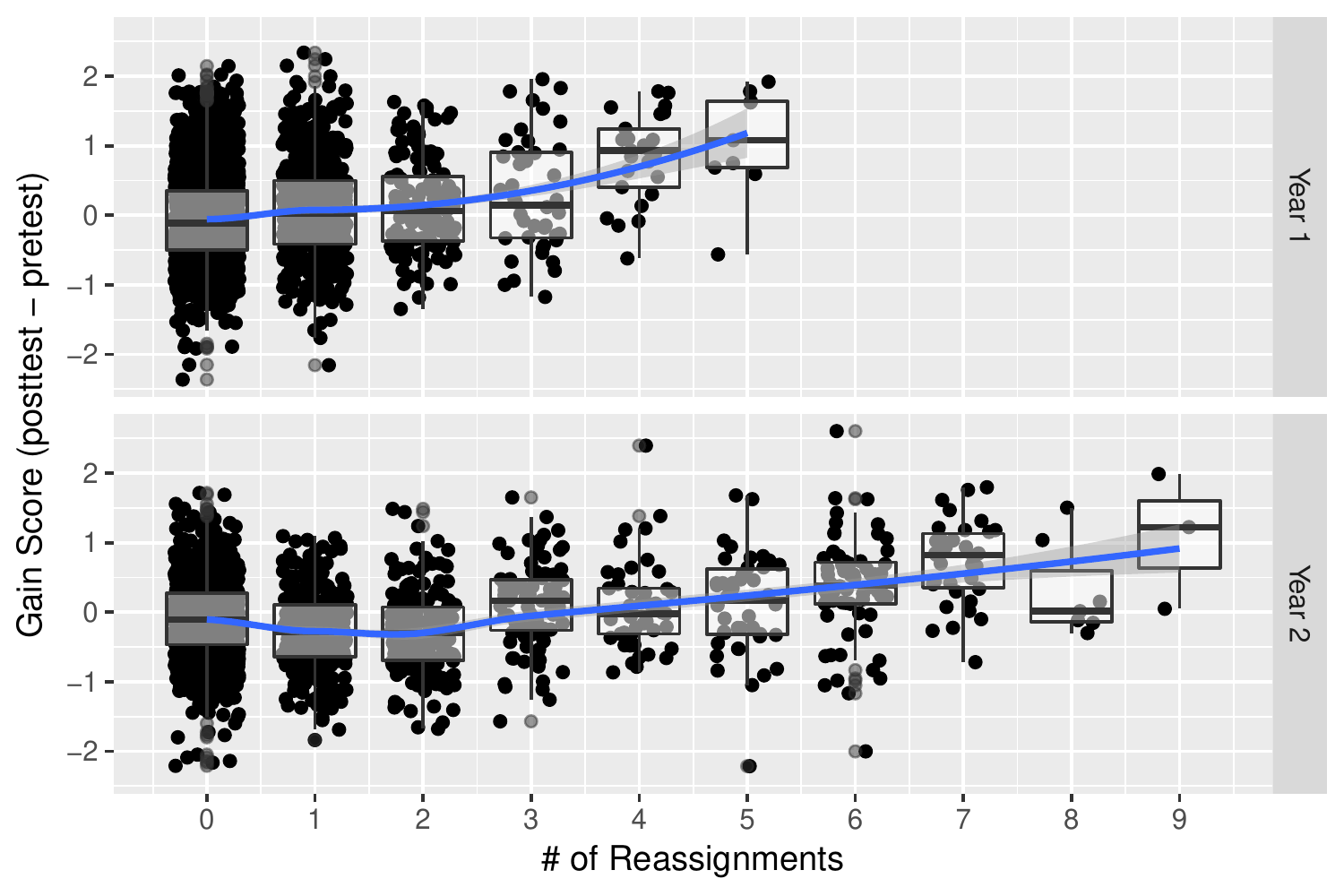}

\caption{Students' Gain scores (posttest minus pre-test) versus the number of times the were reassigned over the course of the year (jittered), with a Loess smoother added.}
\label{fig:cpYyear}
\end{figure}

%\subsection{Average Effects}
Figure \ref{fig:cpYyear} shows students' gain scores---the difference between their posttest and pre-test scores---as a function of the number of times they were reassigned.
(The number of reassignments was jittered---random noise was added on
the horizontal dimension---to avoid overplotting.)
Overall, the relationship between the two variables is positive.
Nevertheless, some non-linearity seems to be present, especially in year 2.
Further, the distribution of the number of reassignments is right-skewed---again, especially in year 2.
Care in modeling the number of reassignments, then, is especially important---observations from students reassigned an unusually large number of times can exert undo influence on a regression model and generate misleading results, particularly in the presence of non-monotonic relationships.
We settled on three different strategies: first, the variable $R^{bin}$ dichotomizes reassignment---$R^{bin}=0$ for students who were never reassigned, and $R^{bin}=1$ for students who were.
Next, $R^{cat}$ defines a categorical variable taking the values $R^{cat}=0,1,2,3,4+$ for students who were reassigned 0, 1, 2, 3 or four or more times, respectively.
Finally, $R^{num}$ is the raw number of reassignments, which we include for completeness.

We used linear models to estimate the effect of reassignment, regressing posttest scores on $R^{bin}$, $R^{cat}$, or $R^{num}$ along with fixed effects for classroom, essentially modeling reassignment as randomly assigned within classroom.
Since this is unlikely to be the case (even approximately), we ran a
second set of models including student level covariates as well:
pretest scores,\footnote{These are measured with error, and missing at a relatively high rate (
15\%). To account for
this, we used regression calibration based on the 20 ``multiple
imputations'' used in the original CTAI study,
\citeN{pane2014effectiveness}.} race, sex, grade, special education
and gifted status, English as a second language (ESL), and free and reduced-price lunch eligibility.

\begin{table}
  \centering
\begin{tabular}{r|c|c|c|}
&\multicolumn{3}{c}{Parametrization}\\
&$R^{bin}$&$R^{cat}$&$R^{num}$\\
&\makecell[c]{Effect of\\ $\ge 1$ Reassignment}&\makecell[l]{Effect of\\  \# Reassignments:}& \makecell[c]{Effect per\\ Reassignment:}\\
\hline
\makecell[r]{No\\Covariates}&-0.14 $\pm$ 0.06 &\makecell[l]{1 :   -0.15 $\pm$ 0.07 \\2 :   -0.15 $\pm$ 0.1 \\3 :   -0.07 $\pm$ 0.15 \\4 +:  -0.12 $\pm$ 0.17 }&-0.04 $\pm$ 0.03 \\
\hline\makecell[r]{Covariate\\ Adjusted}&-0.15 $\pm$ 0.06 &\makecell[l]{1 :  -0.15 $\pm$ 0.07 \\2 :  -0.16 $\pm$ 0.1 \\3 :  -0.08 $\pm$ 0.15 \\4 +:  -0.14 $\pm$ 0.17 }&-0.04 $\pm$ 0.03 \\
\hline
\end{tabular}
\caption{Estimates of effects of reassignment on posttests, with 95\% margins of error.}
\label{effectResults}
\end{table}

The results are reported in Table \ref{effectResults}.
All the effect estimates are negative, indicating that reassigning students may hurt their algebra learning.
The estimated effects decrease in magnitude for three or four reassignments, but these estimates are very noisy---very few students were reassigned more than two times.
The magnitudes of the effects are rather large: \citeN{pane2014effectiveness} reported an effect, in year 2, of about 0.2; the estimated effect of at least one reassignment is 73\% of that.

But what of unmeasured confounding?
For instance, the negative effect may be due to baseline differences
in ability, beyond what is captured in pretest scores.
\citeN{hhh} suggest a method of estimating the sensitivity of a regression to an omitted confounder based on bench-marking from observed confounders.
In order to confound the causal relationship between reassignment and posttests, a confounder would have to predict both.
Roughly speaking, the idea is to widen the confidence interval from an
ostensibly causal linear model to account for the possibility of a hypothetical unmeasured confounder that predicts reassignment and posttests to the same extent as one of the observed covariates.
These ``sensitivity intervals'' account for uncertainty from two sources: random error, and systematic error due to the omission of a confounder.
As is typical, the pretest is our most important measured covariate, both in terms of its prediction of reassignment and of posttest scores.
The sensitivity interval for the effect of being reassigned at least
once on posttest scores, allowing for the possible omission of a
hypothetical confounder at most as important as pretest, is
-0.15 $\pm$
0.14.
This interval is quite wide, implying that such a covariate could
explain much of the observed relationship between reassignment and
posttest scores (or that the relationship may be much stronger).
On the other hand, the sensitivity interval allowing for the possible
omission of a less important hypothetical confounder---one that
predicts posttests as well as ESL status and reassignment as well as
Hispanic ethnicity, the next best observed
predictors---is
-0.15 $\pm$
0.07.
This interval is substantially tighter.
All in all, unmeasured confounding may play an important role here,
but there is good reason to believe that it does not explain all of
the observed relationship.

\begin{figure}
  \centering

\includegraphics[width=\maxwidth]{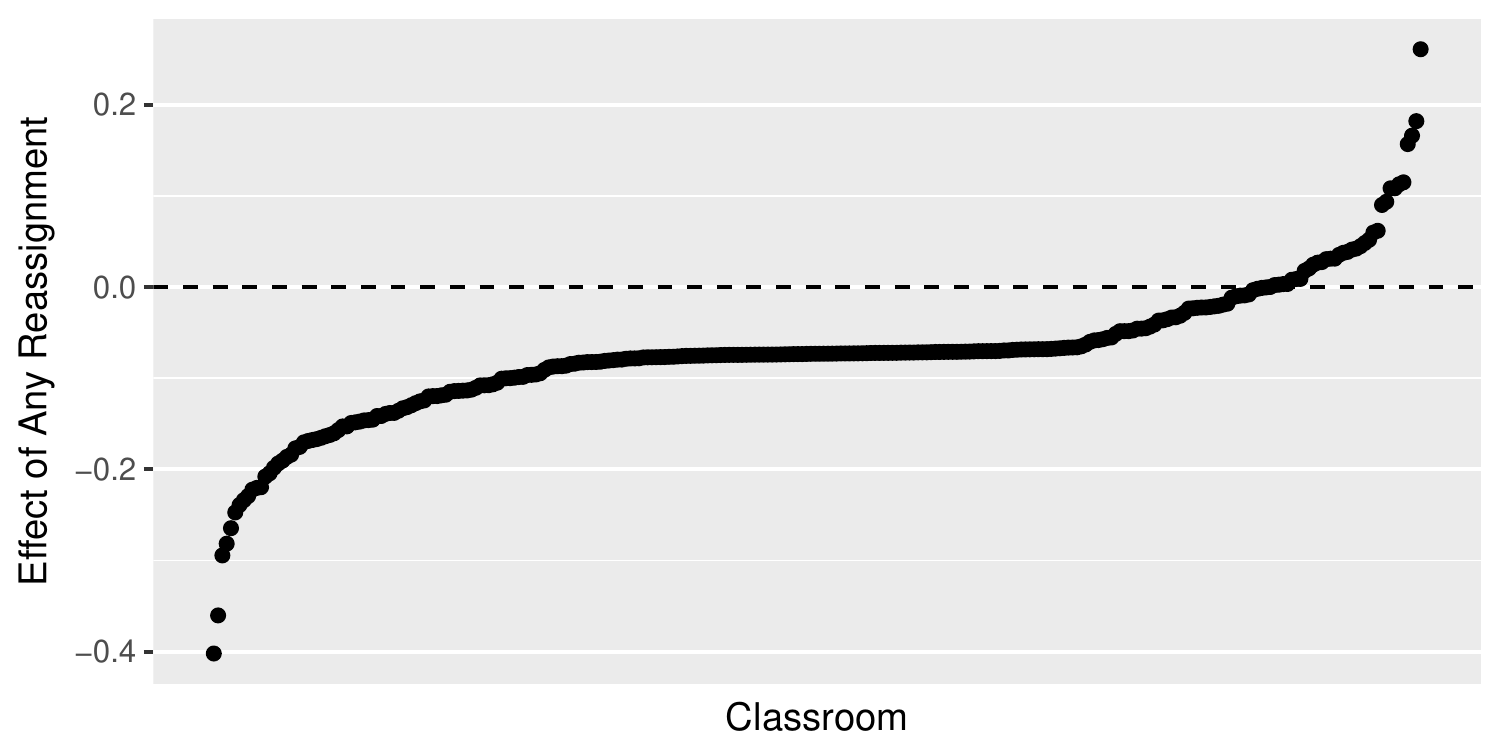}

\caption{The effect of being reassigned at least once, in each
  classroom, as estimated by a multilevel model. The classrooms are
  sorted by these estimated effects. The dotted line indicated an
  effect of zero.}
\label{fig:trtHet}
\end{figure}

The wide variety in the use of reassignment that we have documented
here might suggest that reassignment's treatment effect varies as
well.
Figure \ref{fig:trtHet} shows estimated classroom-specific treatment
effects of $R^{bin}$, being reassigned at least once.
The estimates came from a multilevel model in which posttest scores
were regressed on $R^{bin}$ and student-level covariates, along with
random effects for classroom and random slopes for $R^{bin}$ varying
by classroom.
Unlike fixed effects models, multilevel models ``partially pool'' data
across classrooms to estimate classroom specific effects more precisely \cite{gelmanHill}.
This is especially important given the small sample sizes within
classrooms.
Figure \ref{fig:trtHet} shows a wide variation in the effect of
reassignment across classrooms---the estimated standard deviation of
these effects was
0.18,
larger than the average effect itself.
While the effect was negative in most classrooms, it was positive in some.
This variation could be due to a number of factors, including
differences in the composition of classrooms, but supports the
hypothesis that differences in when and how reassignment is used
lead to differences in its effect.

\section{Summary and Discussion}\label{sec:discussion}

The effectiveness of the Cognitive Tutor software presumably depends
on how much, and how, it is used.
This paper exploited available log data from the high-school arm of the CTAI
effectiveness trial---which yielded an impressive result for the
software in year 2---to describe variation and patterns in the software's
usage, paying particular attention to issues of mastery learning.
We found that:
\begin{itemize}
 \item The amount the software was used varied widely between states
   and decreased, overall, from years 1 to 2.
 \item Year 2 saw the proliferation of ``customized'' curricula in
   three states, altering which units students worked, and in what
   order.
 \item Year 2 saw frequent departures from the standard CTAI unit
   sequence, driven mainly, but not entirely, by customized
   curricula.
 \item Examination of section mastery found that:
 \begin{itemize}
  \item About 85\% of
    worked sections in year 1 and
    81\% in  year 2 ended
    with the student having mastered all of the included skills.
  \item There were three ways students worked a section without
    mastering its contents: exhausting its problems and being
    promoted, being reassigned to a new section by the teacher, and
    ending CT use altogether.
  \item Reassignment was rare, though it was more prevalent in Texas
    and Connecticut than in other states, and more common in year 2
    than 1.
  \item Mastery rates were lower for more advanced curricula.
 \end{itemize}
 \item Examination of reassignment patterns found that:
  \begin{itemize}
   \item Reassignment rates were determined more by factors varying by state and school
     than by classroom.
   \item Reassignment was more common in the second half of the year
     than in the first---particularly in Texas.
   \item Depending on state and year, reassignment typically takes
     place in one of three scenarios: teachers reassigning students
     who have fallen behind, teachers reassigning (almost) the entire
     class together, and teachers assigning (almost) all students
     \emph{out} of a particular section.
   \item Students are most commonly reassigned to the next unit in the
     curriculum---suggesting that teachers may be advancing lagging
     students---but in some cases they might be finely manipulating
     their students' curricula.
   \end{itemize}
 \item Reassignment appears to lower students' performance on the
   post-test relative to that of their peers; however, the effect
   appears to vary widely between classrooms (and, presumably, how
   reassignment is used).
 \end{itemize}

In practice, mastery learning and topic scaffolding often unfolded
quite differently from the vision of CT's designers.
In some cases, the departures were apparently a matter of a student's inability
to achieve mastery quickly enough, and in other cases of educators' preferences that ran counter to CTAI's design.

Notably, both the amount of usage and fidelity to CTAI's design
decreased from years 1 to 2---just as the estimated \emph{effect} of
CTAI increased.
In year 2, when the effect was substantial, students spent less time,
and followed the CTAI curriculum and guidelines less closely than in
year 1, when the estimated effect was negative but statistically
insignificant.
This raises questions as to the roles of structure and mastery in
CTAI's effectiveness.
Does flexibility lead to higher effects? Is mastery learning an
important mechanism for CTAI?
Answers to these questions could prove crucial for optimizing the
realized effectiveness of CTAI and other intelligent tutors.

On the other hand, reassignment appears to hurt students' performance
on the posttest, relative to their classmates (though based on data
from a randomized experiment, this analysis was observational, so
confounding can't be ruled out).
Does this, in contrast, suggest that achieving mastery is an important
component of effective intelligent tutoring?

As educational technology spreads, attention to the details of
implementation may yield important insights---or important
questions---about effectiveness, and the science of when
intelligent tutors work, and when they don't.

\section*{Acknowledgements}
This work was supported by NSF Award \#1420374.

\bibliographystyle{acmtrans}
\bibliography{ct,cp}

\end{document}